\documentclass[twocolumn,showpacs,prc,nofootinbib,floatfix,axodraw]{revtex4}
\usepackage[sort&compress]{natbib}

\usepackage{mathrsfs}
\usepackage{graphicx,booktabs,bm}
\usepackage{overpic}
\usepackage{color}
\usepackage{amssymb}
\usepackage{hyperref}
\usepackage{soul}

\usepackage{mathrsfs,bm,amsmath,amssymb}
\usepackage{longtable,lscape}
\usepackage{txfonts}
\usepackage{amssymb}
\usepackage{indentfirst}
\usepackage{graphicx,booktabs}
\usepackage{multirow,ulem}
\usepackage{subfigure}

\usepackage{float}
\usepackage{array}

\begin{document}
\title{Pion-mass dependence of the nucleon-nucleon interaction}

\author{Qian-Qian Bai}
\affiliation{School of Physics, Beihang University, Beijing 102206, China}

\author{Chun-Xuan Wang}
\affiliation{School of Physics, Beihang University, Beijing 102206, China}

\author{Yang Xiao}
\affiliation{School of Physics, Beihang University, Beijing 102206, China}
\affiliation{Universit\'e Paris-Saclay, CNRS/IN2P3, IJCLab, 91405 Orsay, France}

\author{Li-Sheng Geng}
\email{lisheng.geng@buaa.edu.cn}
\affiliation{School of Physics \&
Beijing Advanced Innovation Center for Big Data-based Precision Medicine, Beihang University, Beijing100191, China.}
\affiliation{School of Physics and Microelectronics, Zhengzhou University, Zhengzhou, Henan 450001, China}

\begin{abstract}
Nucleon-nucleon interactions, both bare and effective, play an important role in our understanding of
the non-perturbative strong interaction, as well as nuclear structure and reactions. In recent years, tremendous
efforts have been seen in the lattice QCD community to derive nucleon-nucleon interactions from first principles. Because of the
daunting computing resources needed, most
of such simulations were still performed with larger than physical light quark masses. In the present work, employing
the recently proposed covariant chiral effective field theory (ChEFT), we study the light quark mass dependence of the nucleon-nucleon
interaction extracted by the HALQCD group. It is shown that the pion-full version of the ChEFT can describe the lattice QCD data
with $m_\pi=469$ MeV and their experimental counterpart reasonably well, while
the pion-less version can describe the lattice QCD data with $m_\pi=672, 837, 1015, 1171$ MeV, for both the $^1S_0$ and $^3S_1$-$^3D_1$ channels. The slightly better description of the single channel than the triplet channel indicates that higher order studies are necessary for the latter. Our results confirmed previous studies that the nucleon-nucleon interaction becomes more attractive for both
the singlet and triplet channels as the pion mass decreases towards its physical value. It  is shown that
 the virtual bound state in the $^1S_0$ channel remains virtual down to the chiral limit, while
 the deuteron only appears for a pion mass smaller than about 400 MeV.  It seems
 that proper chiral extrapolations of nucleon-nucleon interaction are possible for pion masses smaller than
500 MeV, similar to
the mesonic and one-baryon sectors.

\end{abstract}

\date{\today}


\maketitle
\section{Introduction}
Chiral effective field theories have played an important role in our understanding of the non-perturbative strong interaction~\cite{Scherer:2012xha,Bedaque:2002mn,Epelbaum:2008ga,Machleidt:2011zz}. In particular, they have provided a modern theoretical tool to describe nucleon-nucleon interactions, the so-called chiral nuclear forces~\cite{Weinberg:1990rz,Weinberg:1991um}. Nowadays high-precision chiral nuclear forces~\cite{Entem:2003ft,Epelbaum:2004fk,Entem:2017gor,Reinert:2017usi} are widely used as inputs in ab initio methods to study nuclear structure and reactions~\cite{Hammer:2019poc,Epelbaum:2019kcf,RodriguezEntem:2020jgp}.
More recently,  there are increasing interests and efforts in the lattice QCD community to derive nucleon-nucleon (and more generally baryon-baryon) interactions using quark and gluon degrees of freedom~\cite{Aoki:2020bew}.
In addition to providing a non-trivial check on
the underlying theory QCD~\cite{Ishii:2006ec}, potentials derived from lattice QCD simulations can also be used to constrain chiral nuclear forces as well as provide inputs to nuclear structure studies in the unphysical world~\cite{Barnea:2013uqa,McIlroy:2017ssf,Hu:2020djy}.

Lattice QCD simulations have been traditionally performed with larger than physical light quark masses, finite lattice spacing and volume. As a result, multiple extrapolations are needed to obtain physical results. The extrapolation in
quark masses are often referred to as chiral extrapolations. It has been well established in
the mesonic and one-nucleon sectors that chiral extrapolations are reliable at least for relatively small pion masses (for chiral extrapolations of baryon masses and magnetic moments, see, e.g., Refs.~\cite{Ling:2017jyz,Xiao:2018rvd}).
 As for the baryon-baryon sector, the situation is quite different. In the case of chiral extrapolations, very few explicit studies have been performed, and not all of them try to relate simulations performed with different light quark/pion masses~\cite{Barnea:2013uqa,McIlroy:2017ssf,Hu:2020djy}. Though in
the present work we focus on chiral extrapolations, we note that recently finite volume effects in multi-nucleon systems have been
studied~\cite{Eliyahu:2019nkz}.

In Refs.~\cite{Song:2018qqm,Li:2018tbt}, we have studied the $S=-1$ and $S=-2$ hyperon-nucleon and hyperon-hyperon interactions obtained
 in lattice QCD simulations~\cite{Nemura:2017vjc,Miyamoto:2016hqo,Beane:2006gf,Sasaki:2018mzh}. It was shown that the leading order ChEFT can describe lattice QCD data reasonably well, and meanwhile yield results consistent with experimental data. In the present work, we extend the leading order ChEFT to study the nucleon-nucleon interactions
obtained in Ref.~\cite{Inoue:2011ai}. Our purpose is twofold. First, we hope to check whether the covariant ChEFT can describe simultaneously
experimental nucleon-nucleon phase shifts and their lattice QCD counterparts. Though it may seem trivial, this has not been performed in a way similar to corresponding studies in the mesonic or one-baryon sector. Second, for large pion masses, we would like to see whether its pion-less version can succeed in describing the lattice QCD data. It should be noted that ChEFT can also be used
to test the consistency of lattice QCD simulations (see, e.g., Ref.~\cite{Baru:2016evv} for a recent application from such a
 perspective).

This paper is organized as follows. In Sec. II, we briefly introduce the covariant ChEFT approach relevant to the present study. Fitting to the lattice QCD data is performed in Sec. III, followed by discussions, and we summarize in Sec. IV.

\section{Theoretical framework}
The ChEFT we employ in the present work is described in detail in Refs.~\cite{Ren:2016jna,Li:2016mln,Xiao:2018jot,wang:2020myr}. Here
we only provide a concise introduction and relevant formulation needed for chiral extrapolations.

At leading order, the covariant chiral potential in the pion-full version can be written as
\begin{equation}\label{1}
  V_{LO}=V_{CT}+V_{OPE}
\end{equation}
where $V_{CT}$ represents contact contributions and $V_{OPE}$ denotes
one-pion exchanges. The contact contributions are described by
four covariant four-fermion contact terms without derivatives~\cite{Xiao:2018jot}, namely.
\begin{eqnarray}
  V_{CT} &=& C_{S}(\overline{u}(\pmb{p'},s'_{1})u(\pmb{p},s_{1}))(\overline{u}(-\pmb{p'},s'_{2})u(-\pmb{p},s_{2}))\nonumber \\
   &+&  C_{V}(\overline{u}(\pmb{p'},s'_{1})\gamma_{\mu}u(\pmb{p},s_{1}))(\overline{u}(-\pmb{p'},s'_{2})\gamma^{\mu}u(-\pmb{p},s_{2})) \\
   &+&  C_{AV}(\overline{u}(\pmb{p'},s'_{1})\gamma_{5}\gamma_{\mu}u(\pmb{p},s_{1}))(\overline{u}(-\pmb{p'},s'_{2})\gamma_{5}\gamma^{\mu}u(-\pmb{p},s_{2})) \nonumber \\
   &+&   C_{T}(\overline{u}(\pmb{p'},s'_{1})\sigma_{\mu\nu}u(\pmb{p},s_{1}))(\overline{u}(-\pmb{p'},s'_{2})\sigma^{\mu\nu}u(-\pmb{p},s_{2}))\nonumber
\end{eqnarray}
where $\pmb{p}$ and $\pmb{p'}$ are initial and final three momentum, $s_1(s'_1)$, $s_2(s'_2)$ are spin projections, $C_{S,V,AV,T}$ are LECs, and $u(\overline{u})$ are Dirac spinors,
\begin{equation}\label{3}
  u(\pmb{p},s)=N_{P}\begin{pmatrix} 1 \\ \frac{\pmb{\sigma \cdot p}}{E_{P}+M} \end{pmatrix}\chi_{S} \qquad N_{P}=\sqrt{\frac{E_{P}+M}{2M}}
\end{equation}
with $\chi_{S}$ being the Pauli spinor and $E_{P}(M)$ being the nucleon energy (mass). The one-pion-exchange potential in momentum space reads
\begin{eqnarray}
   V_{OPE}(\pmb{p',p}) &=& -(g_{A}^2/4f_{\pi}^{2})\times (\overline{u}(\pmb{p'},s_{1}')\pmb{\tau_{1}}\gamma^{\mu}\gamma_{5}q_{\mu}u(\pmb{p},s_{1}))\nonumber\\
   &\times& (\overline{u}(\pmb{-p'},s_{2}')\pmb{\tau_{2}}\gamma^{\nu}\gamma_{5}q_{\nu}u(\pmb{-p},s_{2}))\nonumber\\
   &/& ((E_{p'}-E_{p})^{2}-(\pmb{p'-p})^{2}-m_{\pi}^{2})
\end{eqnarray}
where $m_{\pi}$ is the pion mass,  $\tau_{1,2}$ are the isospin matrcies, $g_A = 1.26$, and $f_{\pi} = 92.4$ MeV. Note that the leading order covariant potentials already contain all the six spin operators needed to describe nucleon-nucleon scattering.

The contact potentials can be projected into different partial waves in the $|LSJ\rangle$ basis. In the present work, we are only
interested in the $^1S_0$ and $^3S_1$-$^3D_1$ channels. The corresponding partial wave potentials read
\begin{eqnarray}
 V_{1S0}  &=& \xi_{N}[C_{1S0}(1+R^{2}_{\pmb{p}}R^{2}_{\pmb{p'}})+\hat{C}_{1S0}(R^{2}_{\pmb{p}}+R^{2}_{\pmb{p'}})],
\end{eqnarray}
\begin{eqnarray}
 V_{3S1}  &=& \frac{\xi_{N}}{9}[C_{3S1}(9+R^{2}_{\pmb{p}}R^{2}_{\pmb{p'}})+\hat{C}_{3S1}(R^{2}_{\pmb{p}}+R^{2}_{\pmb{p'}})],
\end{eqnarray}

\begin{eqnarray}
  V_{3D1} &=& \frac{8\xi_{N}}{9}C_{3S1}R^{2}_{\pmb{p}}R^{2}_{\pmb{p'}},
\end{eqnarray}

\begin{eqnarray}
 V_{3S1-3D1}  &=& \frac{2\sqrt{2}\xi_{N}}{9}[C_{3S1}R^{2}_{\pmb{p}}R^{2}_{\pmb{p'}}+\hat{C}_{3S1}R^{2}_{\pmb{p}}],
\end{eqnarray}

\begin{eqnarray}
 V_{3D1-3S1}  &=& \frac{2\sqrt{2}\xi_{N}}{9}[C_{3S1}R^{2}_{\pmb{p}}R^{2}_{\pmb{p'}}+\hat{C}_{3S1}R^{2}_{\pmb{p'}}],
\end{eqnarray}
where $ \xi_{N}=4 \pi N_{p}^{2}N_{p}^{'2}$,$R_{\pmb{p}}=|\pmb{p}|/(E_{p}+M)$,$R_{\pmb{p'}}=|\pmb{p'}|/(E_{p'}+M)$. A few remarks are in order. First, compared to the leading order Weinberg approach, there are two LECs for the $^1 S_0$ channel, and two for the $^3 S_1$ channel, instead of only one for
the $^1 S_0$ channel and one for the $^3 S_1$  channel. Second, the two LECs for the  $^3 S_1$ channel are also responsible for the $^3 D_1$ channel and the mixing between $^3 S_1$  and $^3 D_1$. Such a feature enables the rather successful description of the
phase shifts of both$^1 S_0$ and $^3 S_1$ - $^3D_1$ up to laboratory energies of about 300 MeV ~\cite{Ren:2016jna,wang:2020myr}.

The LO contact potentials given above do not contain explicit pion-mass dependent contributions, which are
necessary if one wants to study the pion-mass dependence of lattice QCD nuclear forces. Because $m_\pi^2$ are
counted as of $\mathcal{O}(q^2)$, inclusions of such terms require one go to at least the corresponding order in the momentum expansion.
At such an order, however, the ChEFT has too many LECs~\cite{Xiao:2018jot} which cannot be fixed by the limited lattice QCD data of Ref.~\cite{Inoue:2011ai}. As a result, we only add two pion-mass dependent terms in the $^1S_0$ and $^3S_1$ potentials but keep the momentum expansion at the leading order, similar to, e.g., Ref.~\cite{Haidenbauer:2017dua}. The resulting LECs then read, explicitly,
\begin{equation}
  C_{1S0} \rightarrow C_{1S0}+C_{1S0}^{\pi} m_{\pi}^{2},
\end{equation}
\begin{equation}
 \hat{C}_{1S0} \rightarrow  \hat{C}_{1S0}+\hat{C}_{1S0}^{\pi} m_{\pi}^{2},
\end{equation}
\begin{equation}
  C_{3S1} \rightarrow C_{3S1}+C_{3S1}^{\pi} m_{\pi}^{2},
\end{equation}
\begin{equation}
 \hat{C}_{3S1} \rightarrow  \hat{C}_{3S1}+\hat{C}_{3S1}^{\pi} m_{\pi}^{2}.
\end{equation}

In principle, we could also take into account the fact that in the
one-pion exchange contributions  the LECs, $g_A$ and $f_\pi$, are pion-mass dependent as well, as done, e.g., in
Ref.~\cite{Beane:2002xf}. However, we find that the limited lattice QCD data do not allow one to fully disentangle such contributions from
the contact terms. As a result, in our pion-full theory, we use the physical values for $g_A$ and $f_\pi$. In fact, with
the pion-mass dependence of $f_\pi$ of Ref.~\cite{Beane:2002vs}, we have performed alternative studies of the physical and lattice QCD data with $m_\pi=469$ MeV. Qualitatively similar but quantitatively slightly worse results were obtained. In particular, we note that in the $^3S_1$-$^3D_1$ coupled channel, a better fit of the mixing angle and the $^3D_1$ phase shifts can be achieved if
 the pion mass dependence of $f_\pi$ (and $g_A$) is taken into account, but in this case the description of the
 $^3S_1$ phase shifts deteriorate a bit.

Once the chiral potentials are fixed, we solve the relativistic Kadyshevsky scattering equation~\cite{Kadyshevsky:1967rs} to obtain
the scattering amplitudes,
\begin{eqnarray}
   T(\pmb{p'},\pmb{p}) &=& V(\pmb{p'},\pmb{p})+\int\frac{dp''p''^{2}}{(2\pi)^{3}}V(\pmb{p'},\pmb{p''}) \nonumber\\
   &\times&  \frac{M^{2}_{N}}{2E^{2}_{p''}}\frac{1}{E_{p}-E_{p''}+i\epsilon}T(\pmb{p''},\pmb{p}).
\end{eqnarray}
To avoid ultraviolet divergence, we need to introduce a form factor of the form
$f(\pmb{p},\pmb{p'})=exp[\frac{-\pmb{p}^{2n}-\pmb{p'}^{2n}}{\Lambda^{2n}}]$ with $n=2$ and $\Lambda$ the corresponding cutoff, which is often determined by fitting to data. If the
EFT is properly renormalized, physical results should be independent of the corresponding form factor and the related cutoff.
At this stage, there are still heatly discussions about this issue. For a relevant discussion in the covariant ChEFT, see Ref.~\cite{wang:2020myr}.
In the present work, we simply treat the cutoff $\Lambda$ as a parameter and let its value determined by data.

\section{Results and discussions}
In Ref.~\cite{Inoue:2011ai}, nucleon-nucleon interactions are simulated with five different light quark masses.
The corresponding pion and nucleon masses are given in Table \ref{tab:lat}. The pion mass ranges from 469 MeV to 1171 MeV and
 the corresponding nucleon mass changes from 1161 MeV to 2274 MeV. Clearly for large light quark masses, the ChEFT with explicit pion exchanges is not supposed to work, because as the pion mass increases, its range becomes shorter and it is harder to be distinguished  from the contributions of contact terms. Though it is difficult to pin down precisely where one should replace the pion-full theory with
 the pion-less theory, we choose a pion mass of about 500 MeV.~\footnote{Such a choice is
  also supported by explicit studies playing with both options.} In other words, we will study the lattice QCD data with $m_\pi=469$ MeV by the pion-full theory and those with larger pion masses, $m_\pi=672$, 837, 1015, and 1171 MeV by the pion-less version, where in practice we simply
 turn off the one-pion exchange contributions.

\renewcommand{\arraystretch}{1.5}

\begin{table}[htpb]
\centering
\caption{Pion and nucleon masses (in units of MeV) of lattice QCD simulations~\cite{Inoue:2011ai} studied in
the present work. }\label{tab:lat}
\begin{tabular}{m{1.3cm}<{\centering}| m{1.3cm}<{\centering}|m{1.3cm}<{\centering} | m{1.3cm}<{\centering} | m{1.3cm}<{\centering} |m{1.0cm}<{\centering}}
 \hline\hline
 $m_\pi$  & 469 & 672 & 837 & 1015 & {1171}\\ [0.3ex]
 \hline
$M_{N}$  & 1161  & 1484 & 1749 & 2031 & 2274 \\[0.3ex]
 \hline\hline
\end{tabular}
\end{table}

\subsection{$^1 S_0$ in the pion-full theory}

First we focus on the $^1 S_0$ channel.To study the feasibility of proper chiral extrapolations in the nucleon-nucleon sector, we first fit to
the experimental phase shifts and the lattice QCD ones obtained
with $m_\pi=469$ MeV. We realized that the experimental data and the lattice QCD data alone are not enough
to fix all the four LECs and the cutoff. A very good fit can already be obtained with only three of them, namely,
$C_{1S0}$, $C_{1S0}^\pi$, and the cutoff. The best fitting results are shown in Fig. 1 and the corresponding LECs and cutoff are given in Table II. To see the sensitivity of the results on the cutoff, we have allowed the cutoff to vary by as large as 50 MeV from
the optimal value of 680 MeV. The corresponding results are shown by the blue and grey bands. Clearly, the cutoff dependence is relatively mild as expected.

\begin{figure}[htpb]
\includegraphics[scale=0.4]{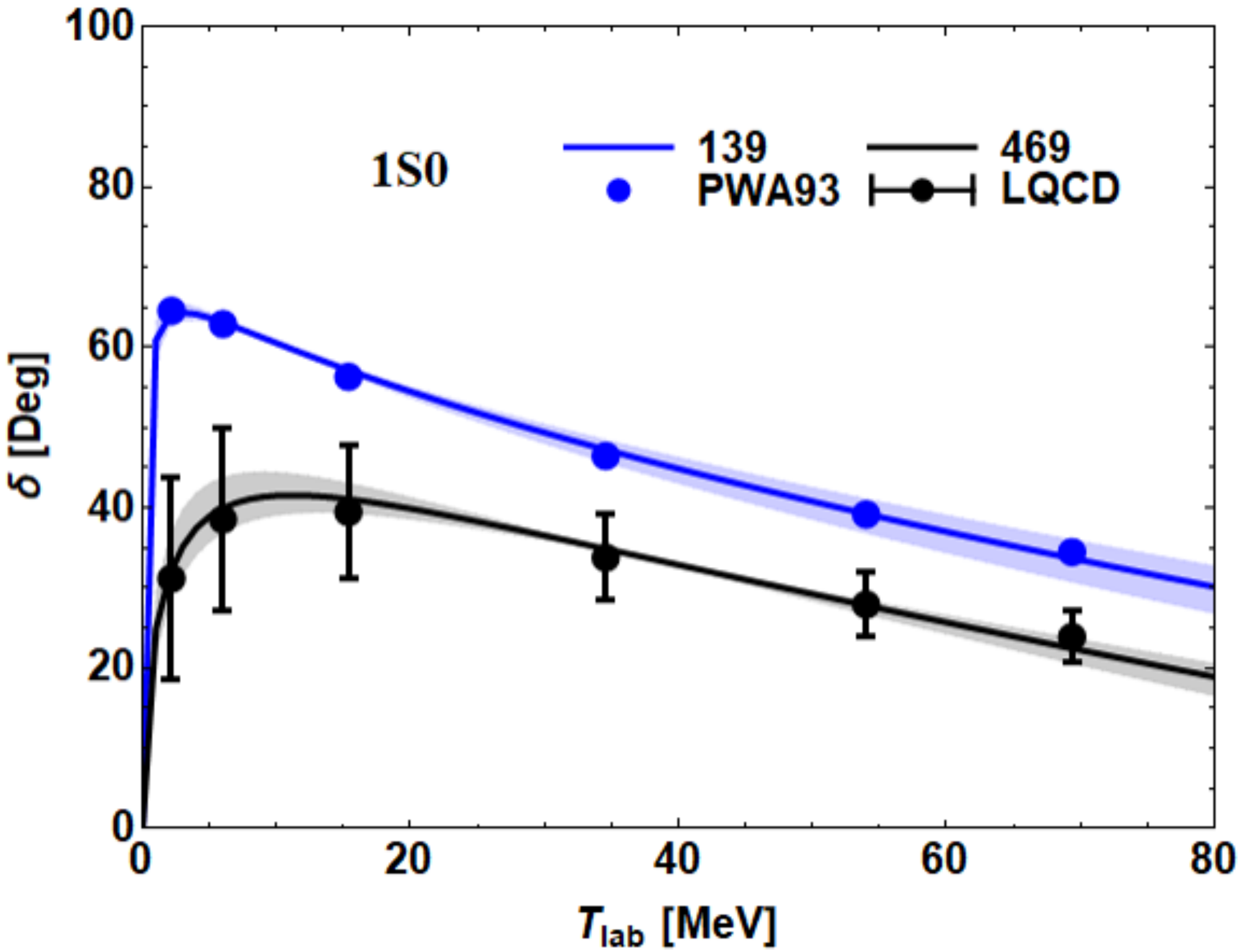}
\caption{$^1 S_0$ phase shifts of the Nijmegen analysis and lattice QCD simulations with $m_\pi=469$ MeV in
comparison with the covariant ChEFT fits.  The blue and grey bands are generated by varying the cutoff from
the optimal value of 680 MeV by $\pm50$ MeV.}\label{fig:1s0ph}
\end{figure}

\begin{table}[htpb]
\caption{ Values of the LECs of the best fit to the Nijmegen analysis and lattice QCD simulations with $m_\pi$ = 469 MeV for the $^1 S_0$ channel.}\label{tab:1s0ph}
\centering
\begin{tabular}{m{2.7cm}<{\centering}| m{2.7cm}<{\centering}|m{2.7cm}<{\centering}}
 \hline\hline
   $C_{1S0}$ [MeV$^{-2}$] &  $C_{1S0}^\pi$ [MeV$^{-4}$] & $\Lambda$ [MeV] \\ [0.3ex]
 \hline
   0.046 &  0.119E-06 & 680\\[0.3ex]
 \hline\hline
\end{tabular}
\end{table}

Clearly, the covariant ChEFT can provide a quite good description of both the experimental data and lattice QCD data simultaneously. This suggests that if more lattice QCD data with  pion masses between 469 and 139 MeV are available, one should be able to perform reliable extrapolations to
the physical point, similar to the cases of mesonic and one-baryon sectors. Such a finding is indeed very encouraging. In addition, the cutoff value of the best fit, about 680 MeV, looks very reasonable. As we will see, in the  $^3 S_1$ - $^3 D_1 $ channel things are a bit different.

\subsection{$^3 S_1$ - $^3 D_1 $ in the pion-full theory}

Now we turn to the $^3 S_1$ - $^3 D_1 $ channel. As we already briefly mentioned in Sec. II, the covariant ChEFT can describe the $^3 S_1$ - $^3 D_1 $  channel already at leading order, though at a price, i.e., the same two LECs are responsible for the $^3 S_1$,$^3 D_1$, and their mixing angle. At the physical point, these two LECs are able to describe the Nijmegen phase shifts reasonably well~\cite{Ren:2016jna,wang:2020myr}. The situation is a bit different in the present case. First, we notice that the best fit to  the physical and lattice QCD phase shifts yields a cutoff of about 300 MeV, which is a bit too small given the fact the lattice QCD pion mass is 469 MeV. Therefore, we decide to fix the cutoff to some specified values and study  to what extent one can describe simultaneously the experimental and lattice QCD phase shifts. We choose four cutoff values, 500 MeV, 1000 MeV, 1500 MeV, and 2000 MeV for such a purpose.
Furthermore, for the triplet channel case, we only fit to those lattice QCD simulations of $T_\mathrm{lab.}<40$ MeV, different from the singlet channel case, as it is impossible to obtain a good fit to all the lattice QCD data. This fitting strategy will also be employed for
the pion-less case in the following subsection.

\begin{figure*}[htpb]
	\centering
	\includegraphics[scale=0.27]{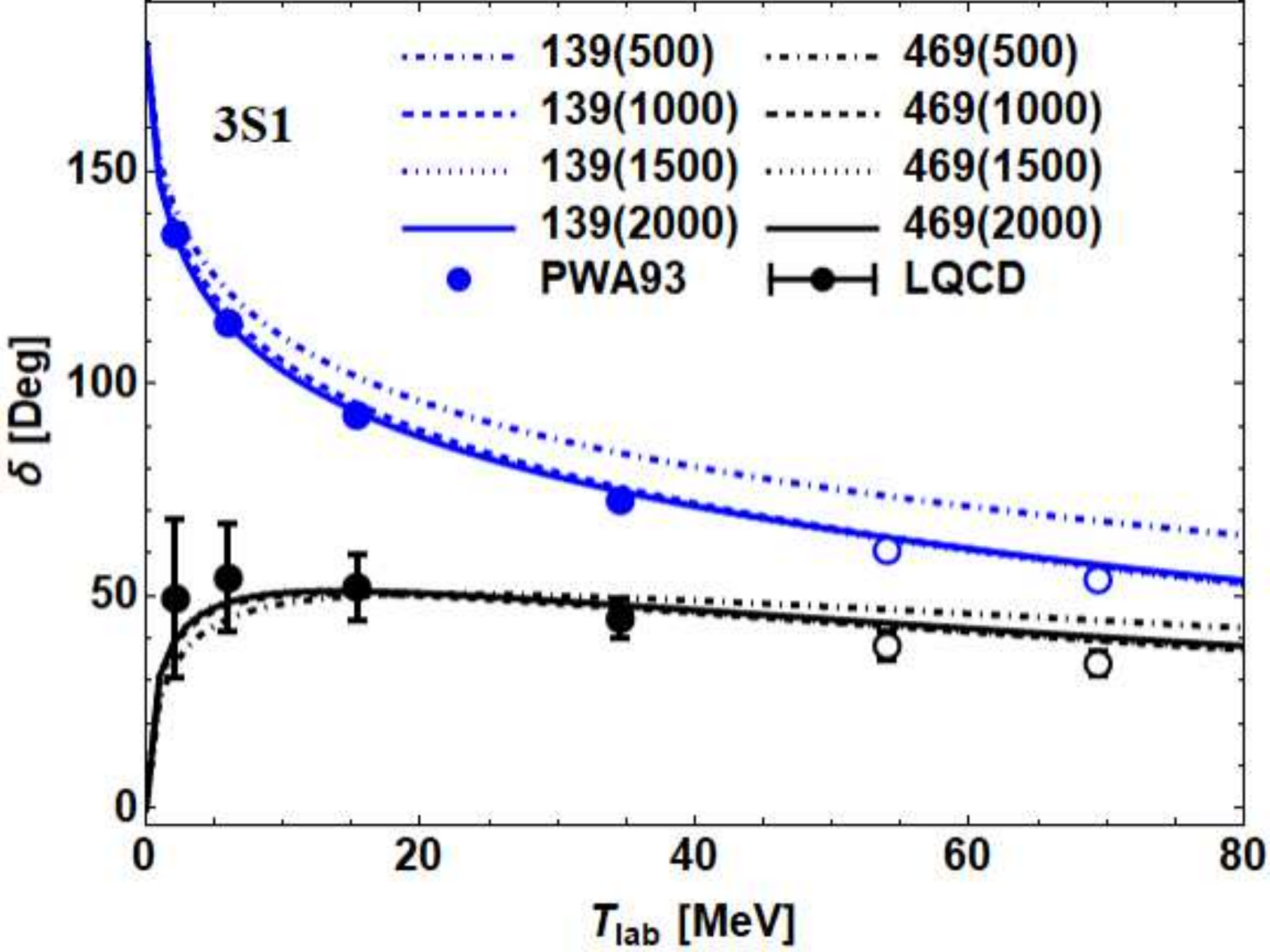}
    \includegraphics[scale=0.267]{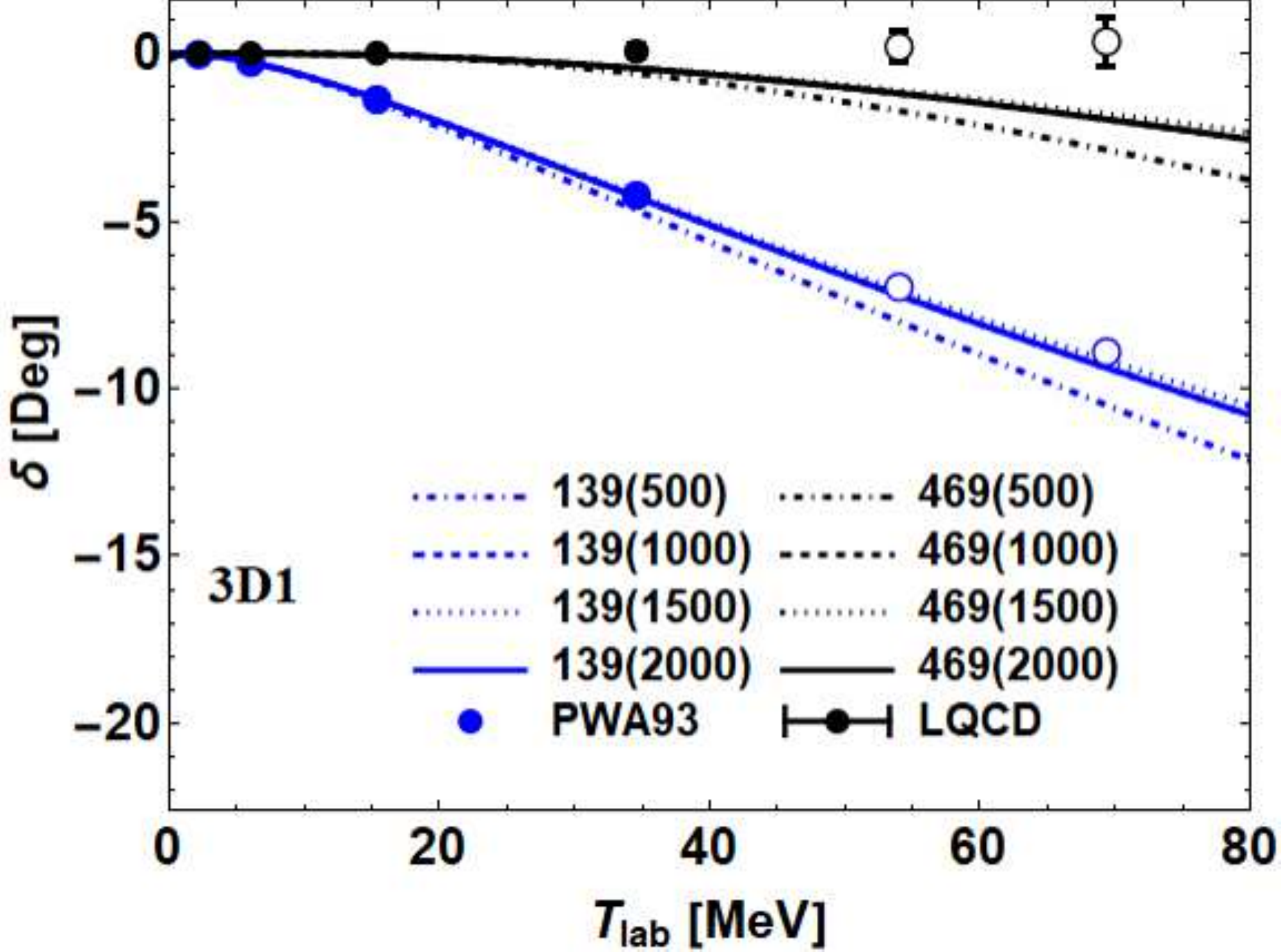}
    \includegraphics[scale=0.265]{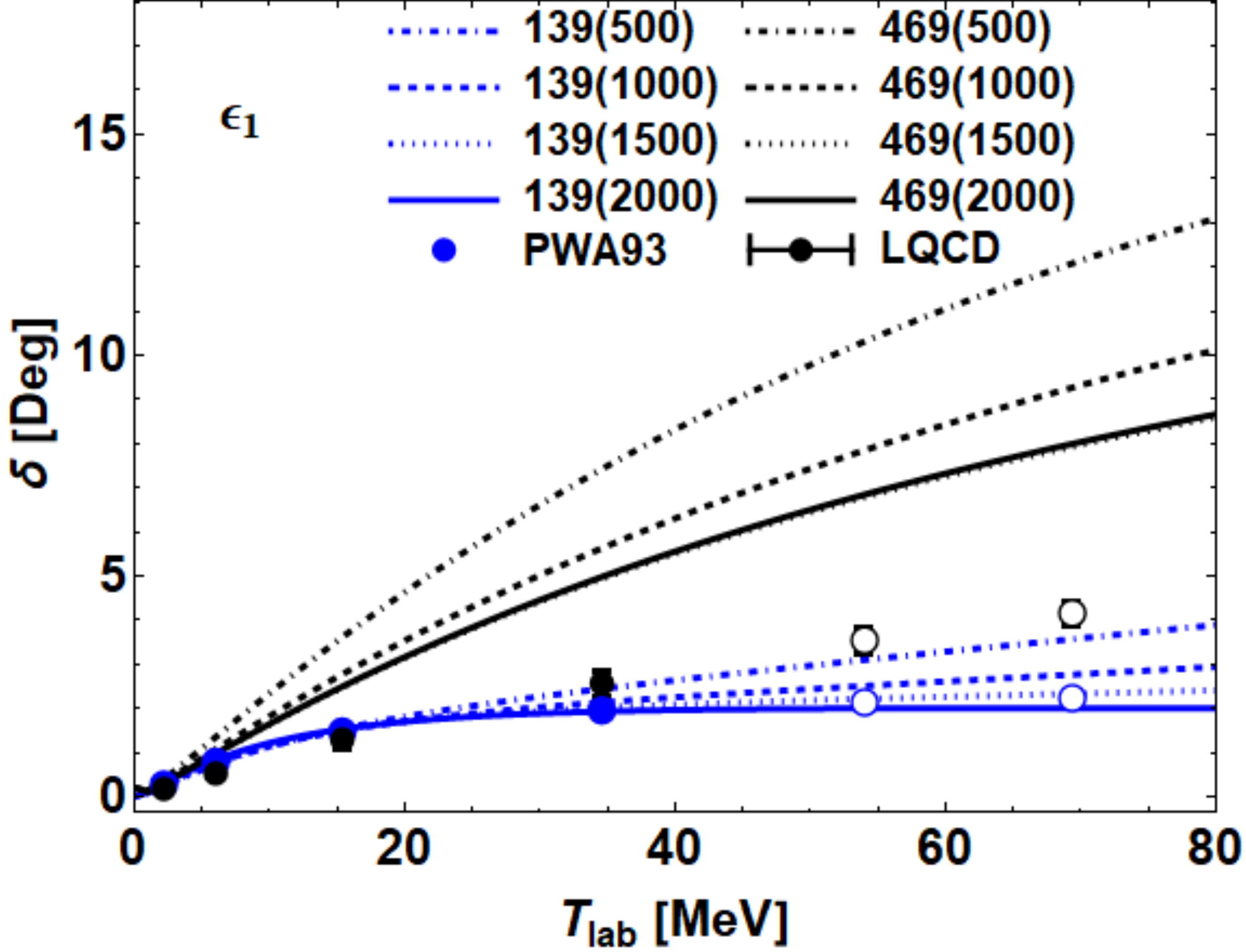}
	\caption{$^3 S_1 -^3 D_1$ phase shifts of the Nijmegen analysis and lattice QCD simulations with $m_\pi=469$ MeV in
    comparison with the covariant ChEFT fits. The solid blue/black circles represent experimental/lattice data fitted, while those empty
circles are not included in the fits.}\label{fig:3s1ph}
\end{figure*}

The results are shown in Fig.~\ref{fig:3s1ph}. It seems that for larger cutoff values, the $^3 S_1$ and $^3 D_1$ phase shifts can be described relatively well, at least up to  $T_\mathrm{lab.}\approx $ 40 MeV . On the other hand, the lattice QCD mixing angle $\epsilon_1$ stays very close to its experimental counterpart, which cannot be fully reproduced by the leading order covariant ChEFT. Though this could be an artifact due to the strong constraint that the same two LECs are responsible for the three observables, the relative insensitivity of $\epsilon_1$ on the light quark masses needs to be better understood
in the future. Another interesting observation is that as the pion mass increases, the magnitude of both the
$^3 S_1$ and the $^3 D_1$ phase shifts decreases, to such a point that for $m_{\pi}$ = 469  MeV the deuteron becomes unbound, as already noticed in
Ref~\cite{Inoue:2011ai}.

 A few remarks are in order concerning the EFT description of the singlet and triplet channels. At the physical point, the covariant ChEFT can describe the singlet and triplet channels quite well at least up to $T_\mathrm{lab.}\approx50$ MeV with a single cutoff of about 750 MeV~\cite{Ren:2016jna}. While when the lattice QCD data with $m_\pi=469$ MeV are fitted together with the physical data, one cannot describe both channels simultaneously. The best fits yield a cutoff of 680 MeV for the singlet channel and about 300 MeV for the triplet channel, indicating an inconsistency between the lattice QCD data and the LO ChEFT.~\footnote{As mentioned
 in the introduction, similar inconsistencies between the NPLQCD determinations of the scattering length and effective range of $^1S_0$ and $^3S_1$ at $m_\pi=450$ MeV using the effective range approximation and those obtained with low energy theorems
 have been noted in Ref.~\cite{Baru:2016evv}.} In addition, we note that the rather similar mixing angle $\epsilon_1$ at the
physical and unphysical pion masses remains to be understood from the EFT perspective. Although one definitely should first check these inconsistencies from the EFT perspective by going to higher orders, studies of the same observables at the same unphysical pion masses by lattice QCD groups other than the HALQCD Collaboration are highly welcome. We note that there are still ongoing discussions on the validity of the HALQCD method and the L\"uscher method in the baryon-baryon
sector, particularly for heavy pion masses,  see, e.g., Refs.~\cite{Iritani:2017rlk,Beane:2017edf,Yamazaki:2017gjl}.

\begin{table}[htpb]
\centering
\caption{ Values of the LECs of the various fits with different cutoffs to the Nijmegen analysis and lattice QCD simulations with $m_\pi$ = 469 MeV for the $^3 S_1$ - $^3 D_1$ channel. }\label{tab:3s1ph}
\begin{tabular}{m{2.7cm}<{\centering}| m{2.7cm}<{\centering}|m{2.7cm}<{\centering}}
 \hline\hline
   $C_{3S1}$ [MeV$^{-2}$] &  $C_{3S1}^\pi$ [MeV$^{-4}$] & $\Lambda$ [MeV]\\ [0.3ex]
 \hline
   0.761E-04   & -0.189E-09 & 500\\[0.3ex]
  \hline
   -0.815E-04 &  0.609E-11 & 1000\\[0.3ex]
    \hline
   -0.118E-02 &  -0.375E-08& 1500\\[0.3ex]
    \hline
   -0.327E-02 & -0.267E-07 & 2000\\[0.3ex]
 \hline\hline
\end{tabular}
\end{table}
\subsection{$^1S_0$ and $^3S_1$ scattering lengths}
\begin{figure*}[htpb]
\includegraphics[scale=0.3]{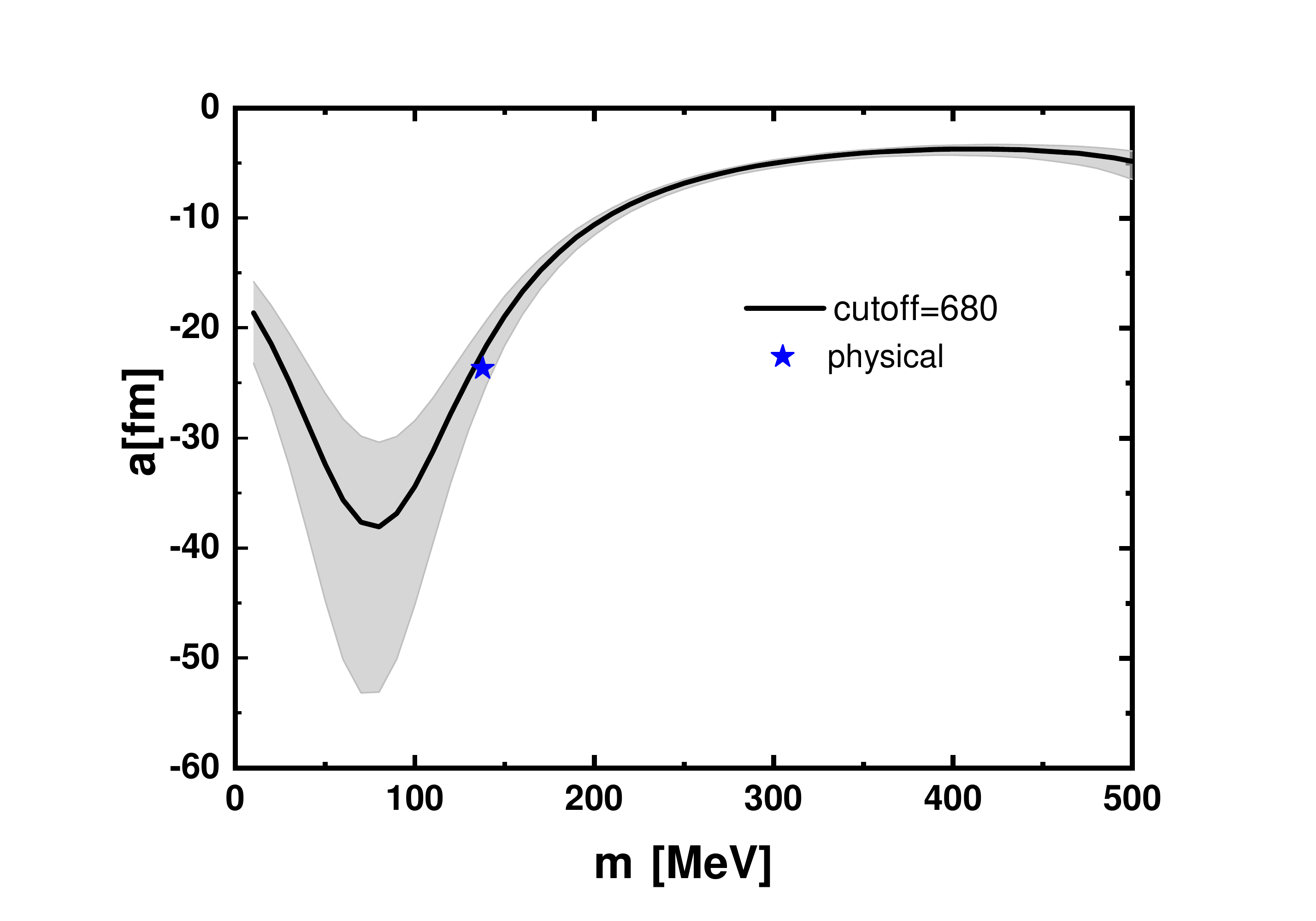}
\includegraphics[scale=0.292]{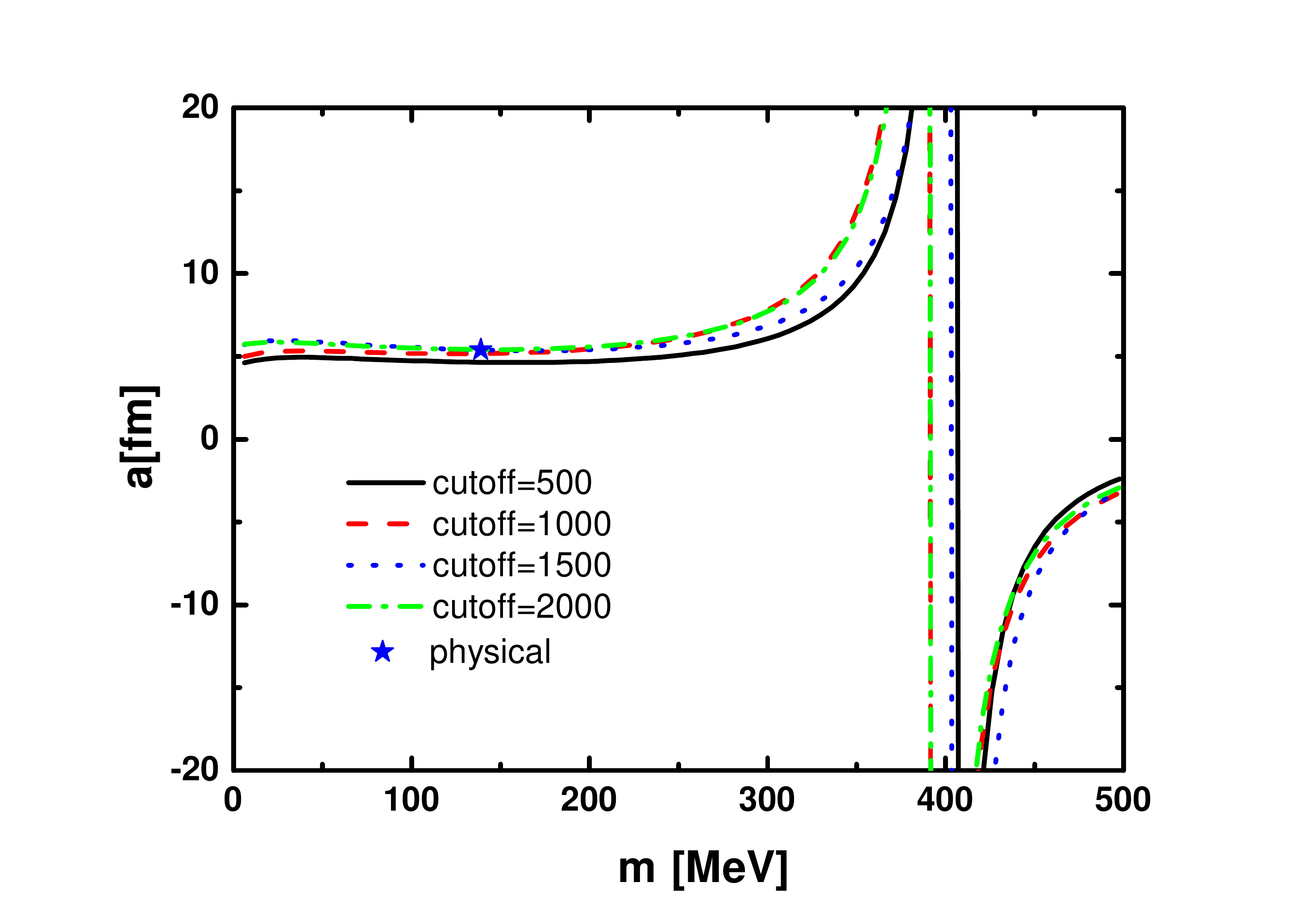}
\caption{ $^1 S_0$ (left) and $^3S_1$ (right) scattering lengths as functions of the pion mass in comparison with
the experimental data, $a_{1S0}=-23.7$ fm and $a_{3S1}=5.4$ fm, denoted by blue stars. The  grey band is generated by varying the cutoff from
the optimal value of 680 MeV by $\pm50$ MeV.}\label{fig:scat}
\end{figure*}

Once the relevant LECs are determined in the pion-full theory, we can study the evolution of physical observables as a function of
the pion mass.  In Fig.~\ref{fig:scat}, we show the scattering lengths for the $^1S_0$ and $^3S_1$ channels as functions of the
pion mass with the LECs tabulated in Tables \ref{tab:1s0ph} and \ref{tab:3s1ph}.~\footnote{
 The dependence of the nucleon mass on the pion mass is described using the
covariant chiral perturbation theory up to $\mathcal{O}(p^3)$ of Ref.~\cite{Ling:2017jyz}, with the two LECs fixed by
fitting to the experimental nucleon mass and the HALQCD nucleon mass with $m_\pi=469$ MeV. }First, we notice that the experimental data are reproduced quite well, though they are not explicitly fitted, reflecting that the descriptions of experimental phase shifts close to threshold are
reasonably good. Second, for the $^1S_0$ channel, though the interaction becomes more attractive for $m_\pi$ smaller than its physical value,
it does not become strong enough to generate a bound state. In addition, the interaction strength reaches the largest around $m_\pi\approx 80$ MeV and then decreases toward the chiral limit. On the other hand, for the $^3S_1$ channel, as expected, a bound state appears somewhere
around $m_\pi=400$ MeV, as indicated by the ``threshold-like'' behavior (see, e.g. Ref.~\cite{Zhou:2014ila}).  It would be interesting to
 compare the present predictions with future lattice QCD data to deepen our understanding of the strong nuclear force, particularly, its dependence on the light quark masses.

We note by passing that though there have been a number of previous studies on
the $^1S_0$ and $^3S_1$ scattering lengths in terms of light quark mass dependences~\cite{Soto:2011tb,Chen:2010yt,Beane:2002xf,Beane:2001bc,Epelbaum:2002gk,Epelbaum:2002gb}, we refrain
from a detailed comparison with their results due to the different inputs used and different strategies taken to
perform the light quark mass evolution. Nonetheless, it is interesting to note that our predicted scattering lengths in
the chiral limit seem to agree with those of Ref.~\cite{Epelbaum:2002gb}, at least qualitatively.
\subsection{$^1S_0$ and $^3S_1$-$^3D_1$ in the pion-less theory }

\begin{figure}[htpb]
\includegraphics[scale=0.2]{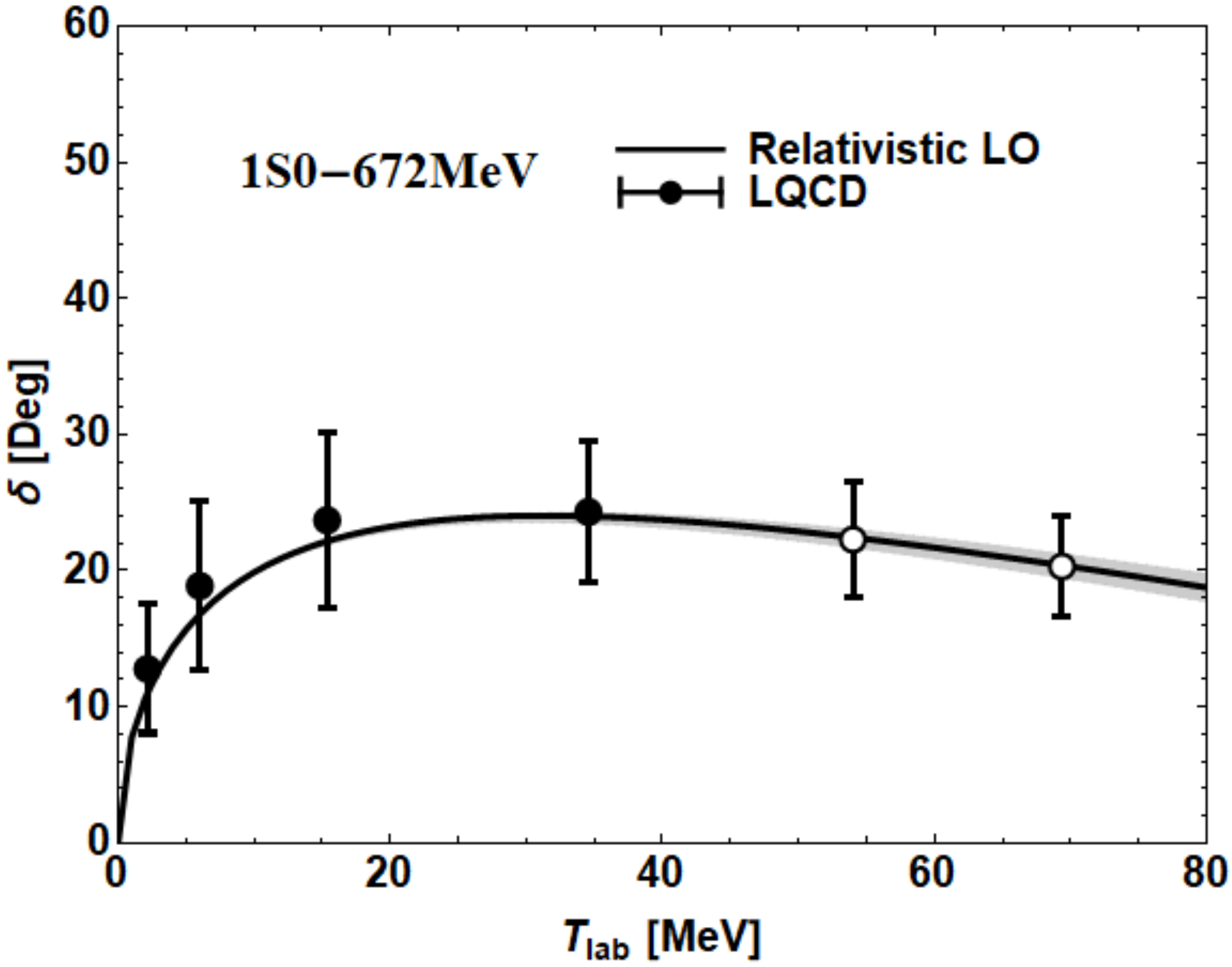}
\includegraphics[scale=0.2]{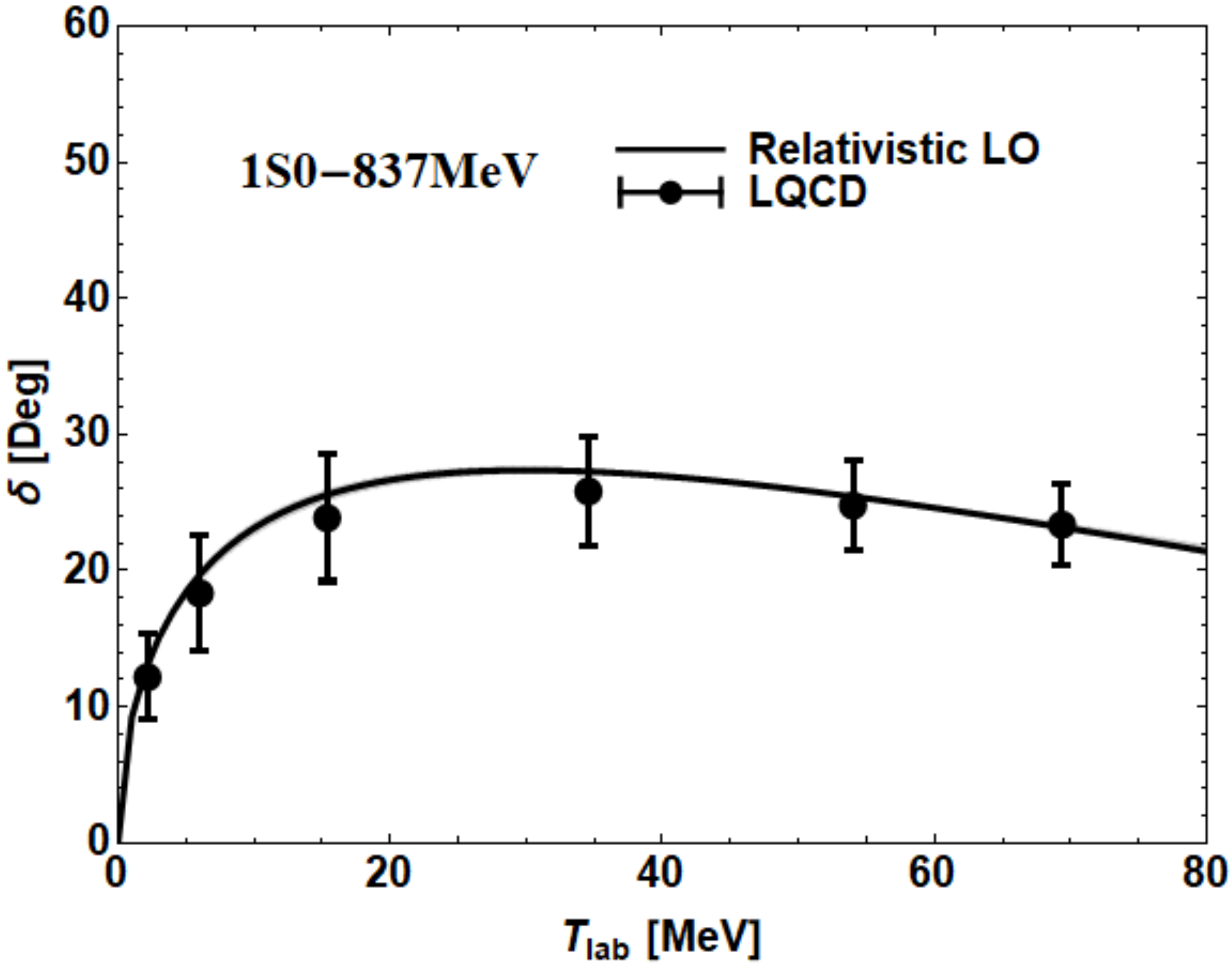}
\includegraphics[scale=0.2]{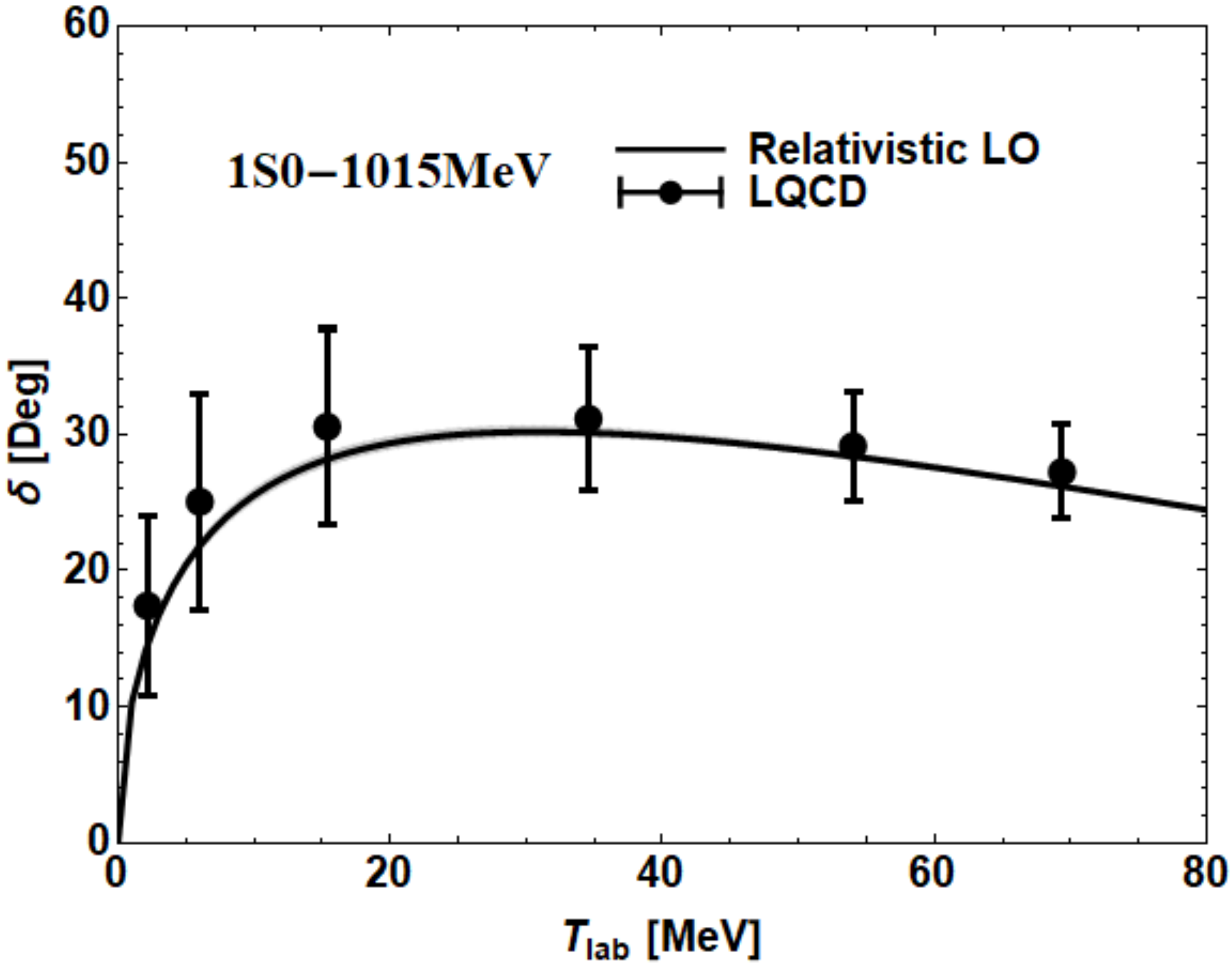}
\includegraphics[scale=0.2]{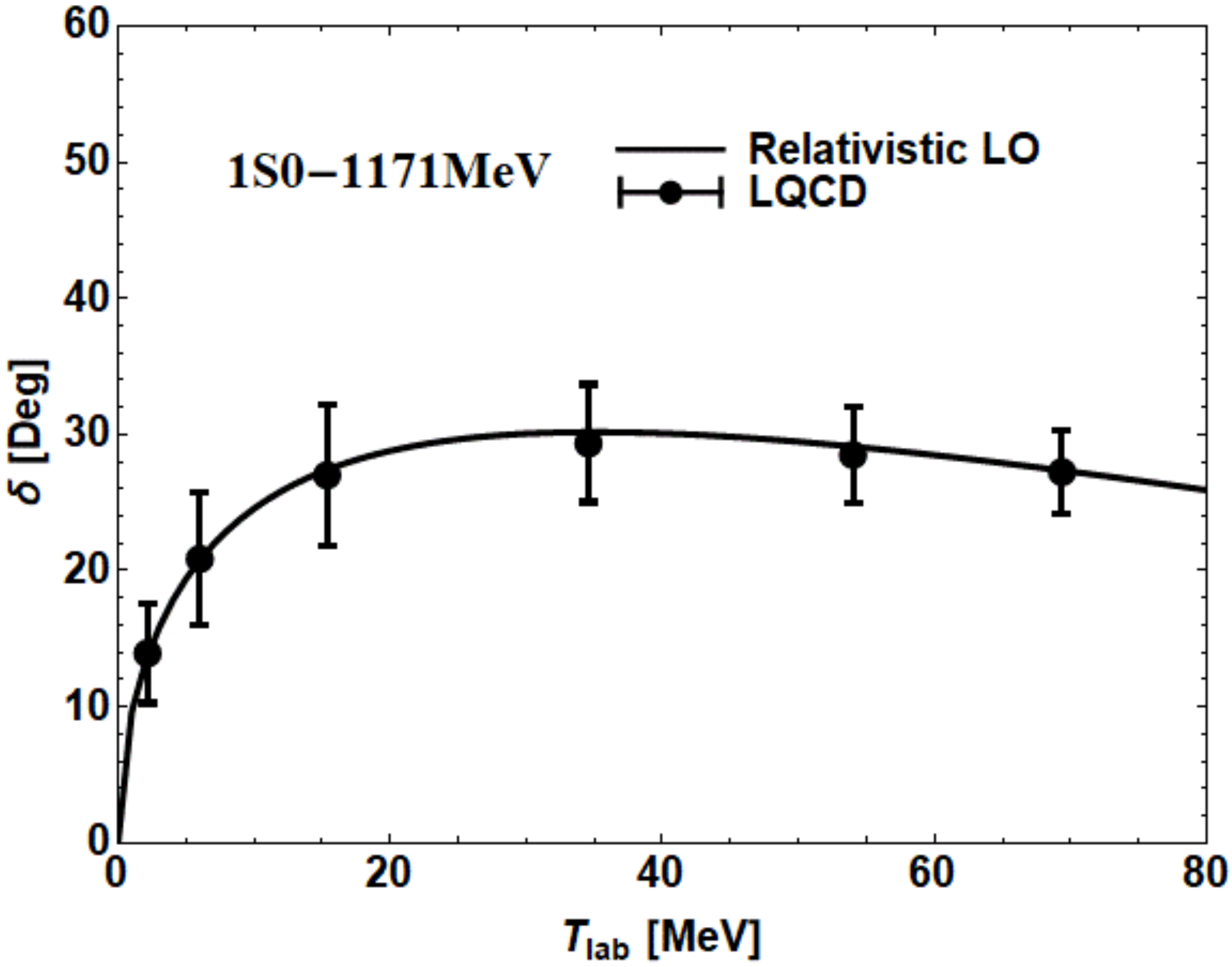}
\caption{ $^1 S_0$ phase shifts of the lattice QCD simulations with $m_\pi$ = 672, 837, 1015, 1171 MeV in
comparison with the covariant ChEFT fits. The (small) grey bands are generated by varying the cutoff from
the optimal value of 730 MeV by $\pm50$ MeV.} \label{1s0lat}
\end{figure}

\begin{figure*}[htpb]
\includegraphics[scale=0.2]{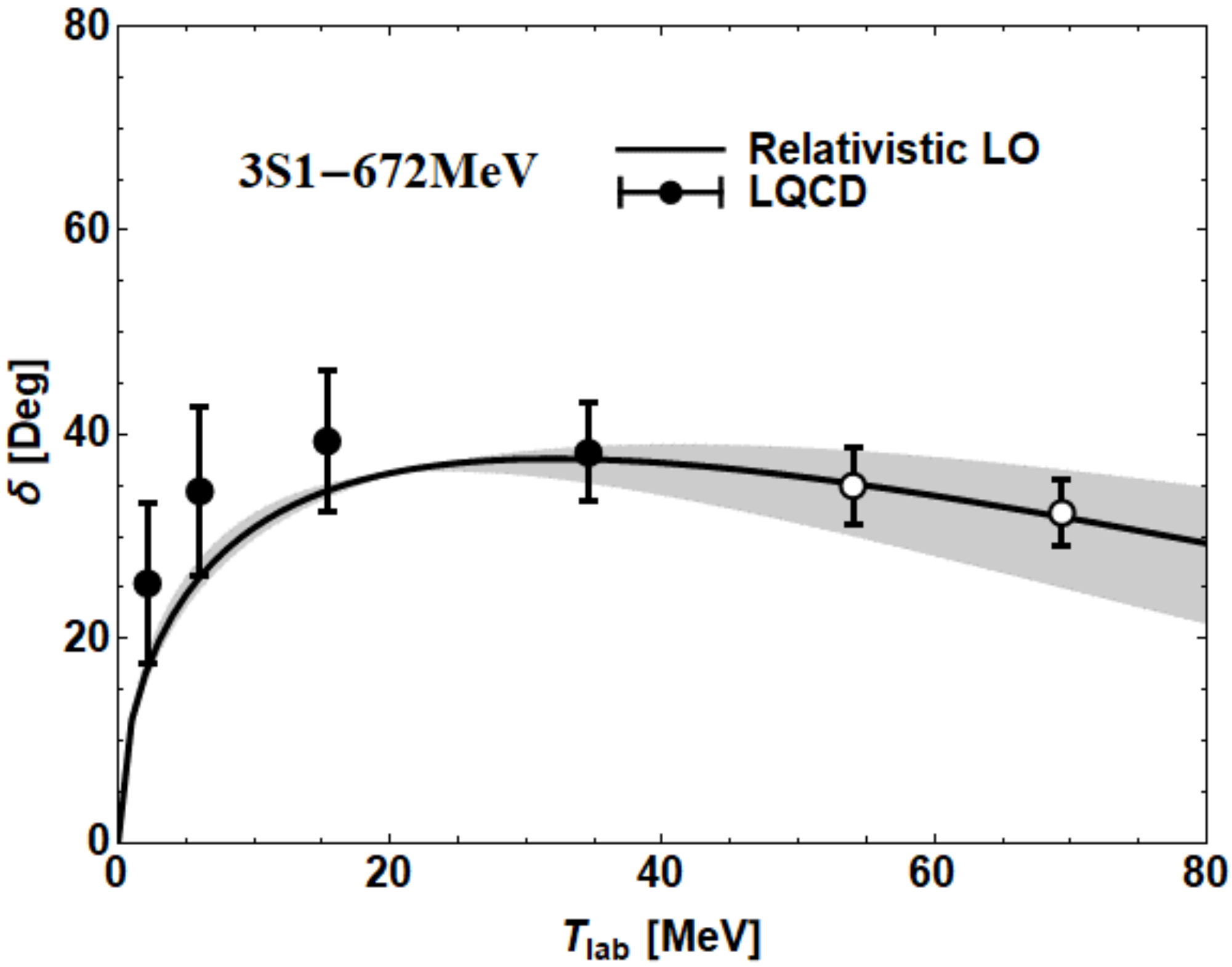}
\includegraphics[scale=0.2]{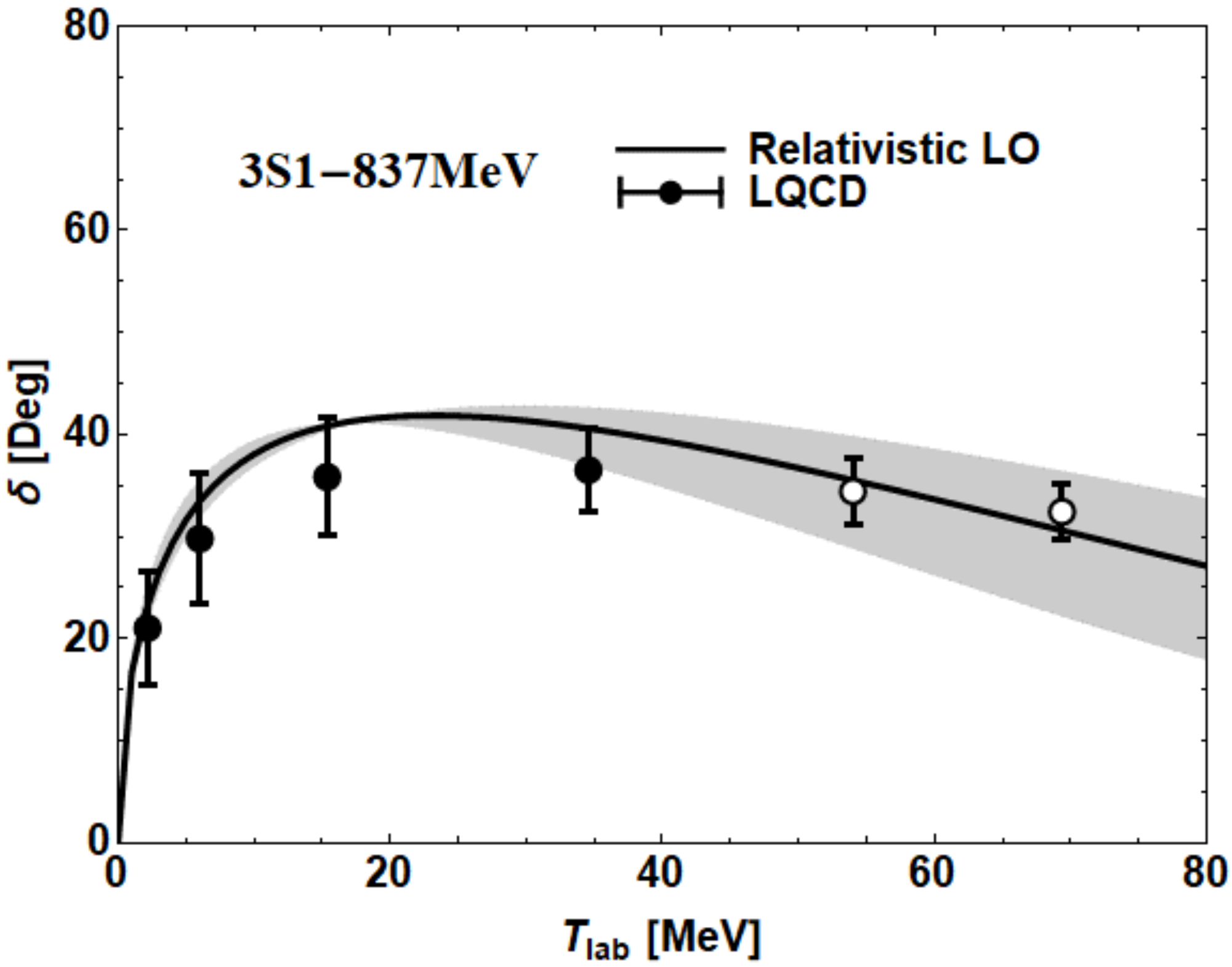}
\includegraphics[scale=0.2]{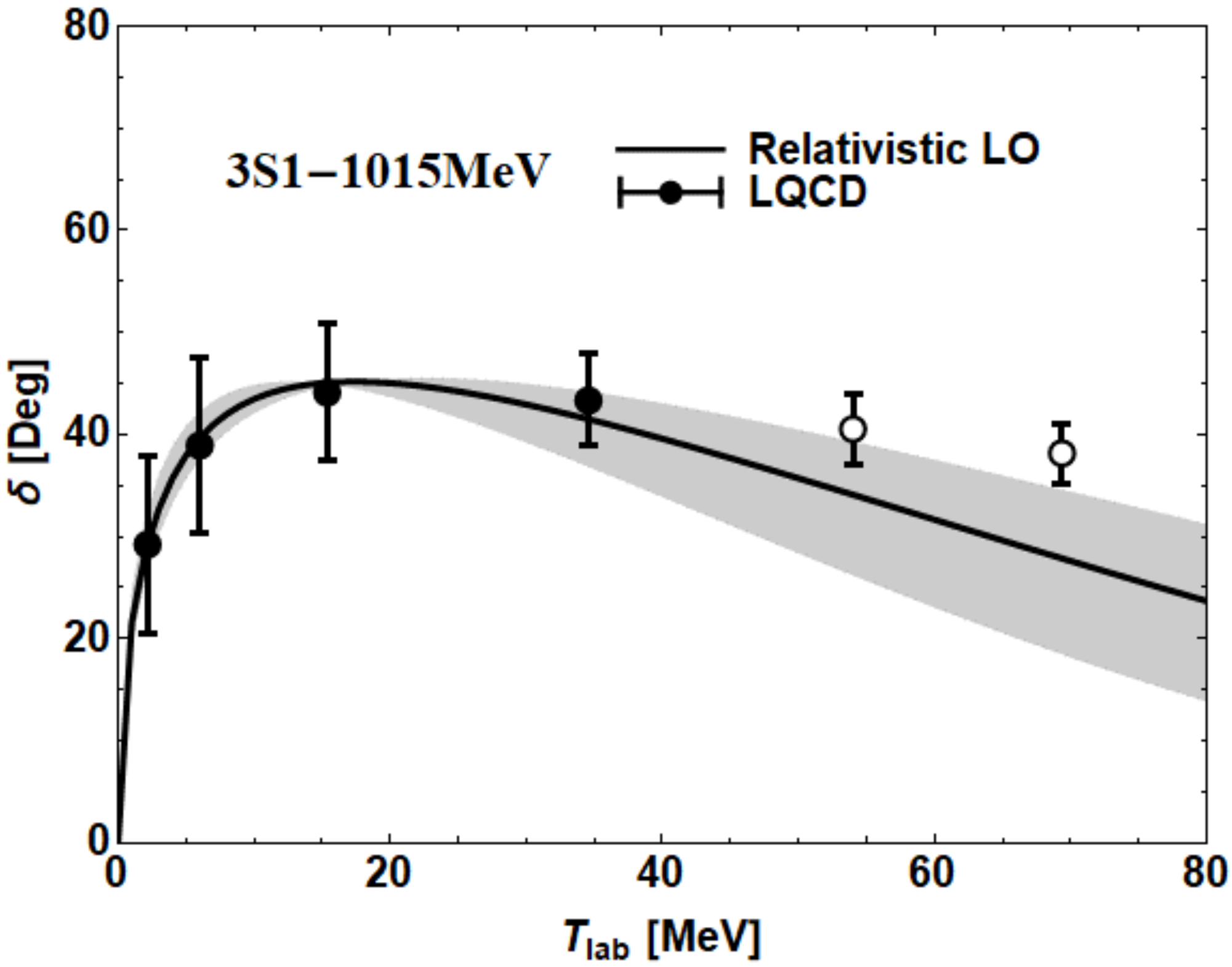}
\includegraphics[scale=0.2]{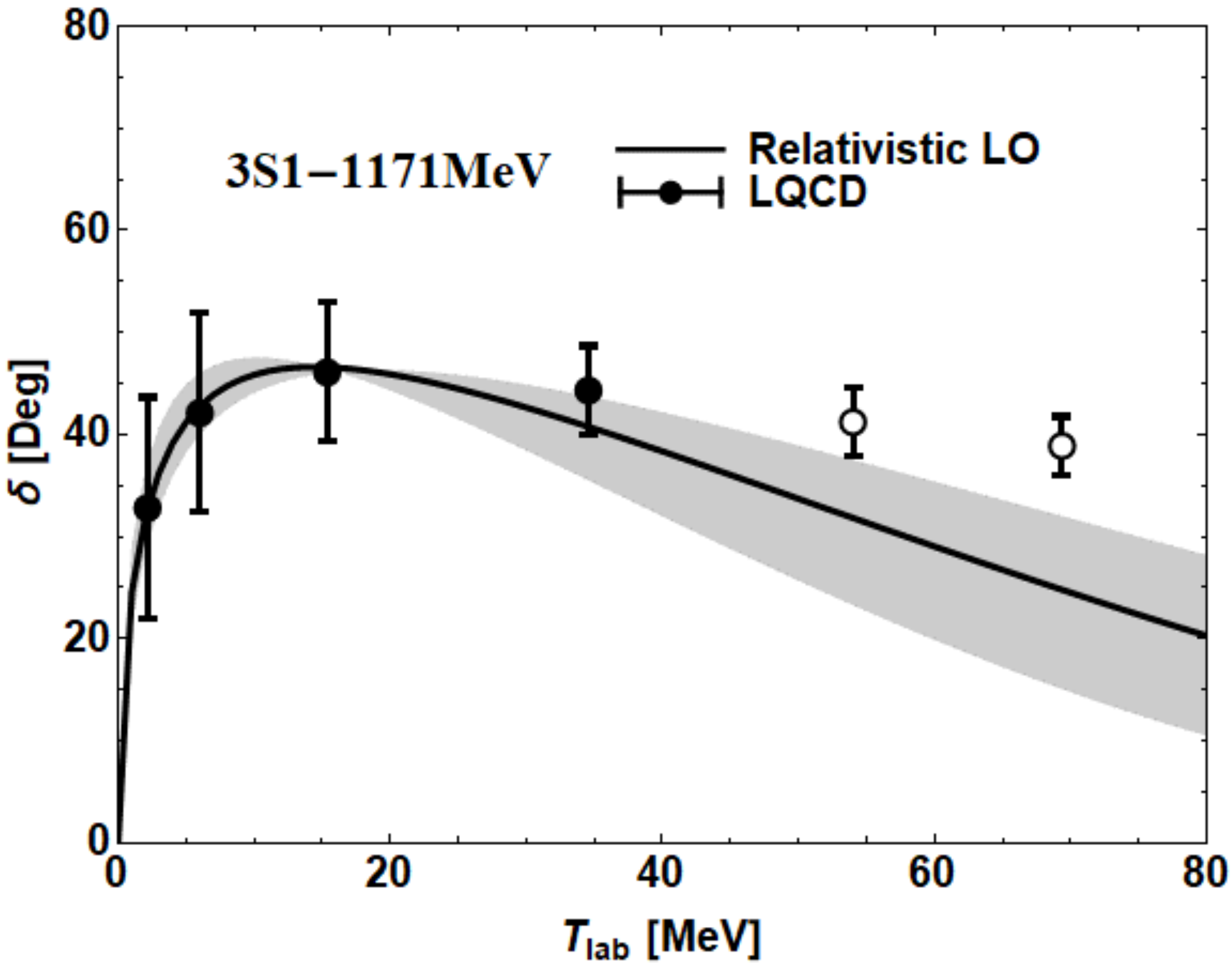}
\includegraphics[scale=0.2]{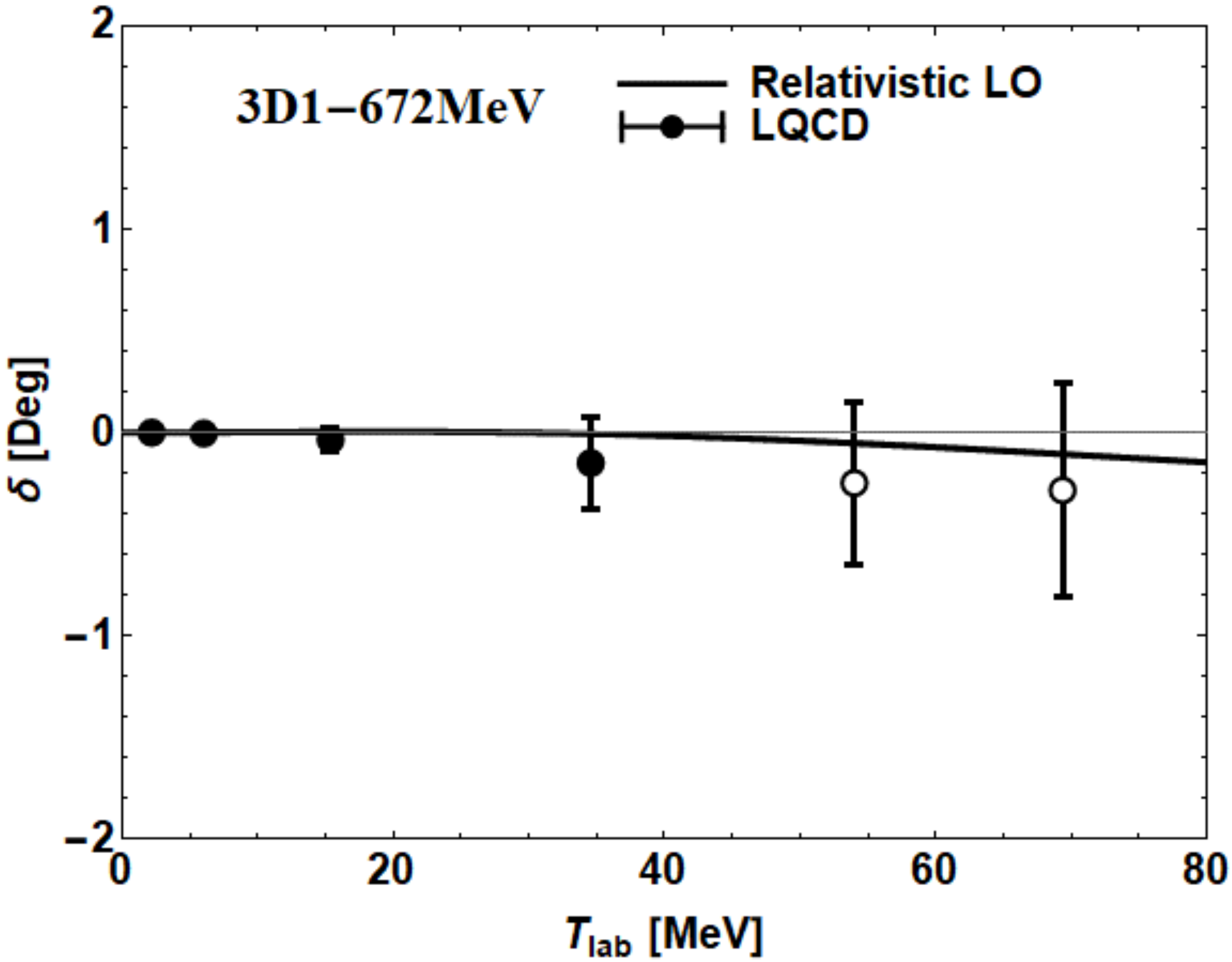}
\includegraphics[scale=0.2]{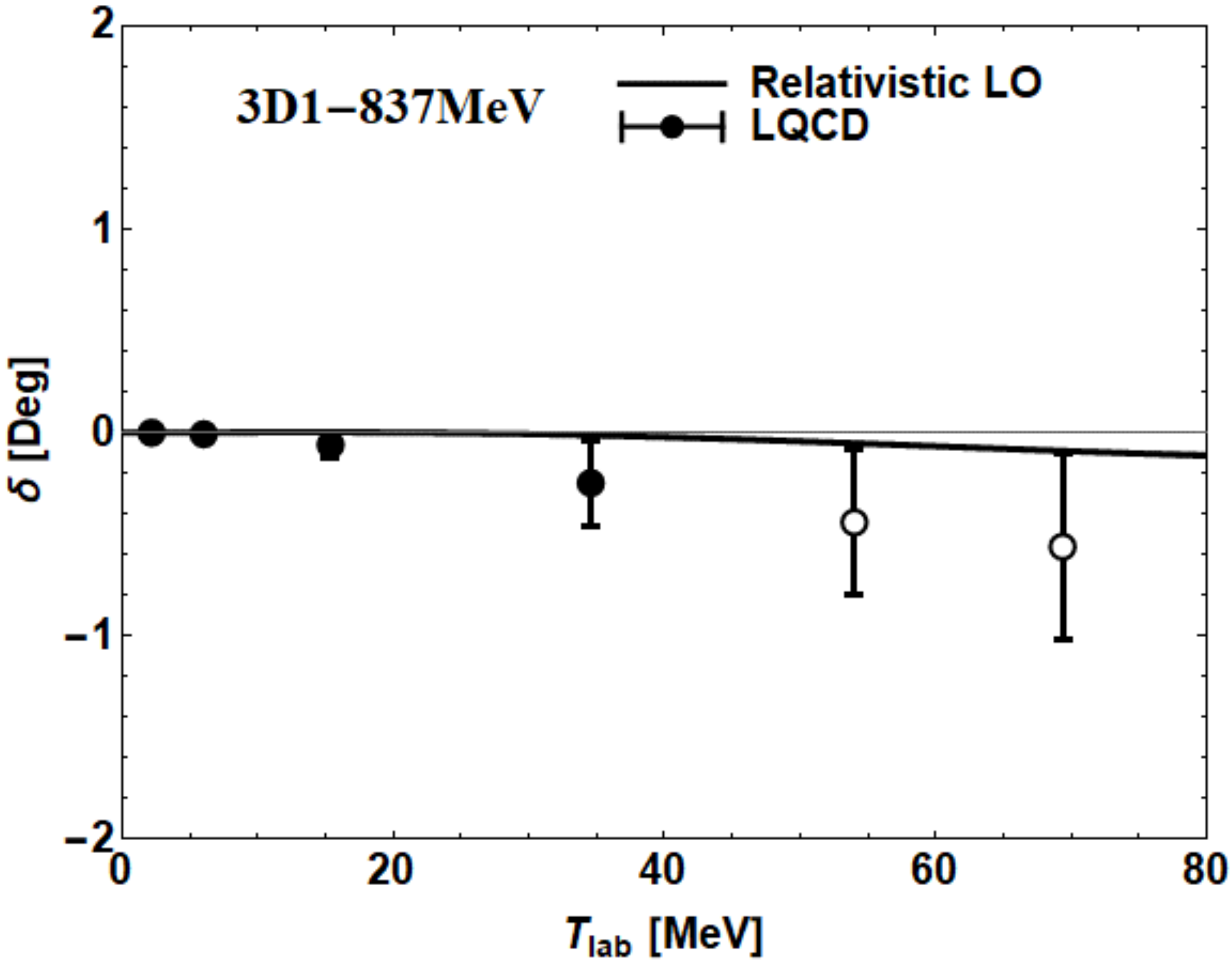}
\includegraphics[scale=0.2]{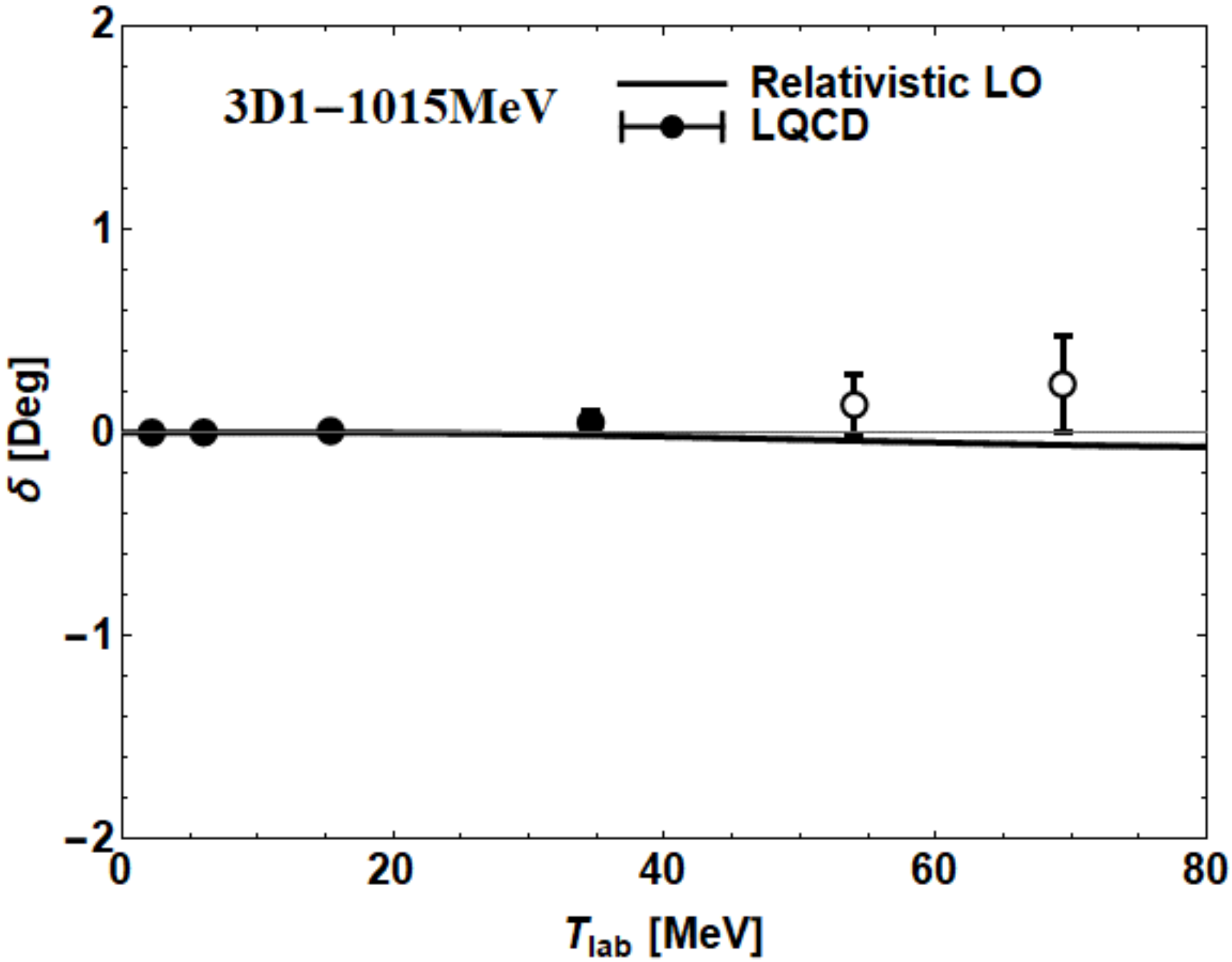}
\includegraphics[scale=0.2]{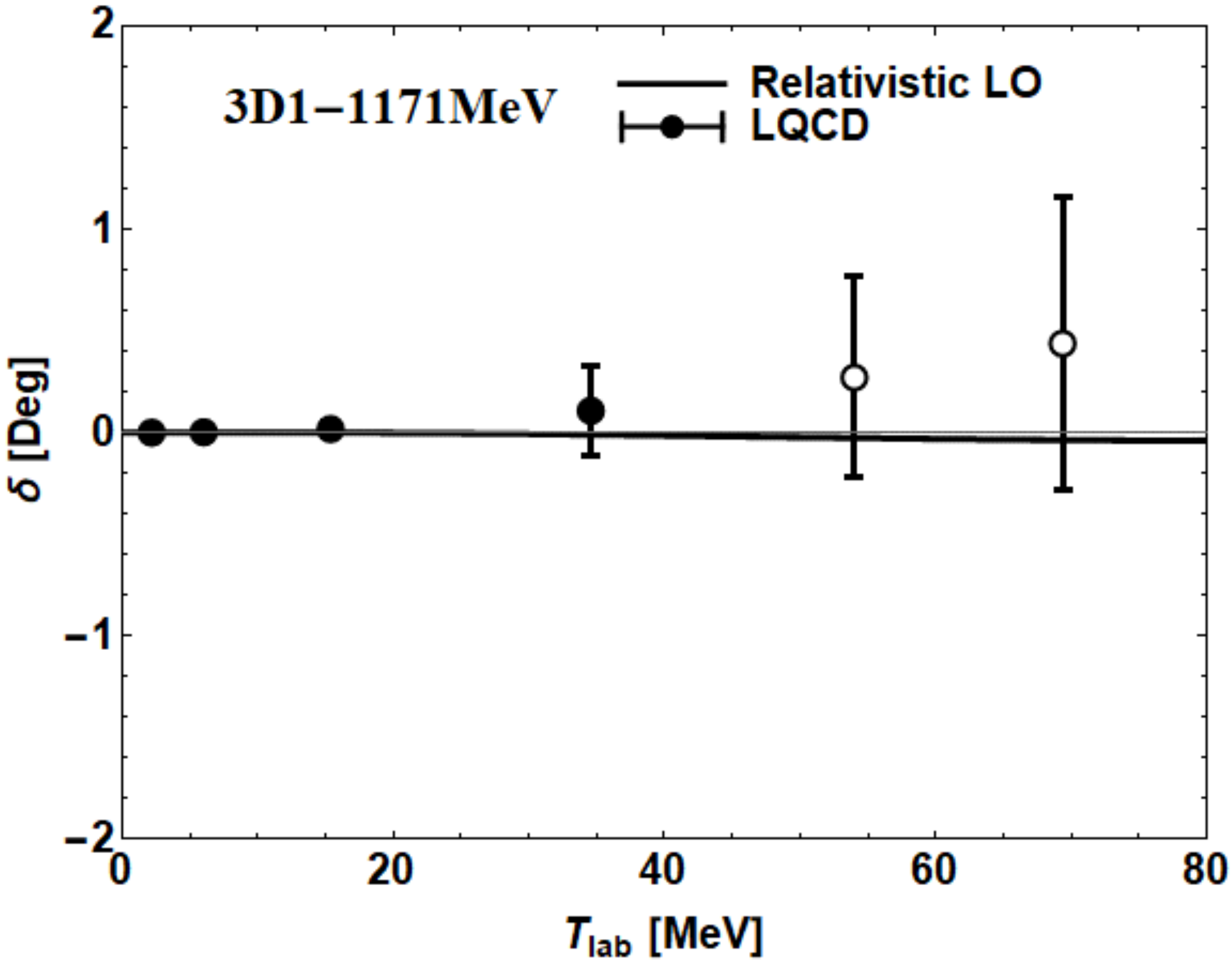}
\includegraphics[scale=0.2]{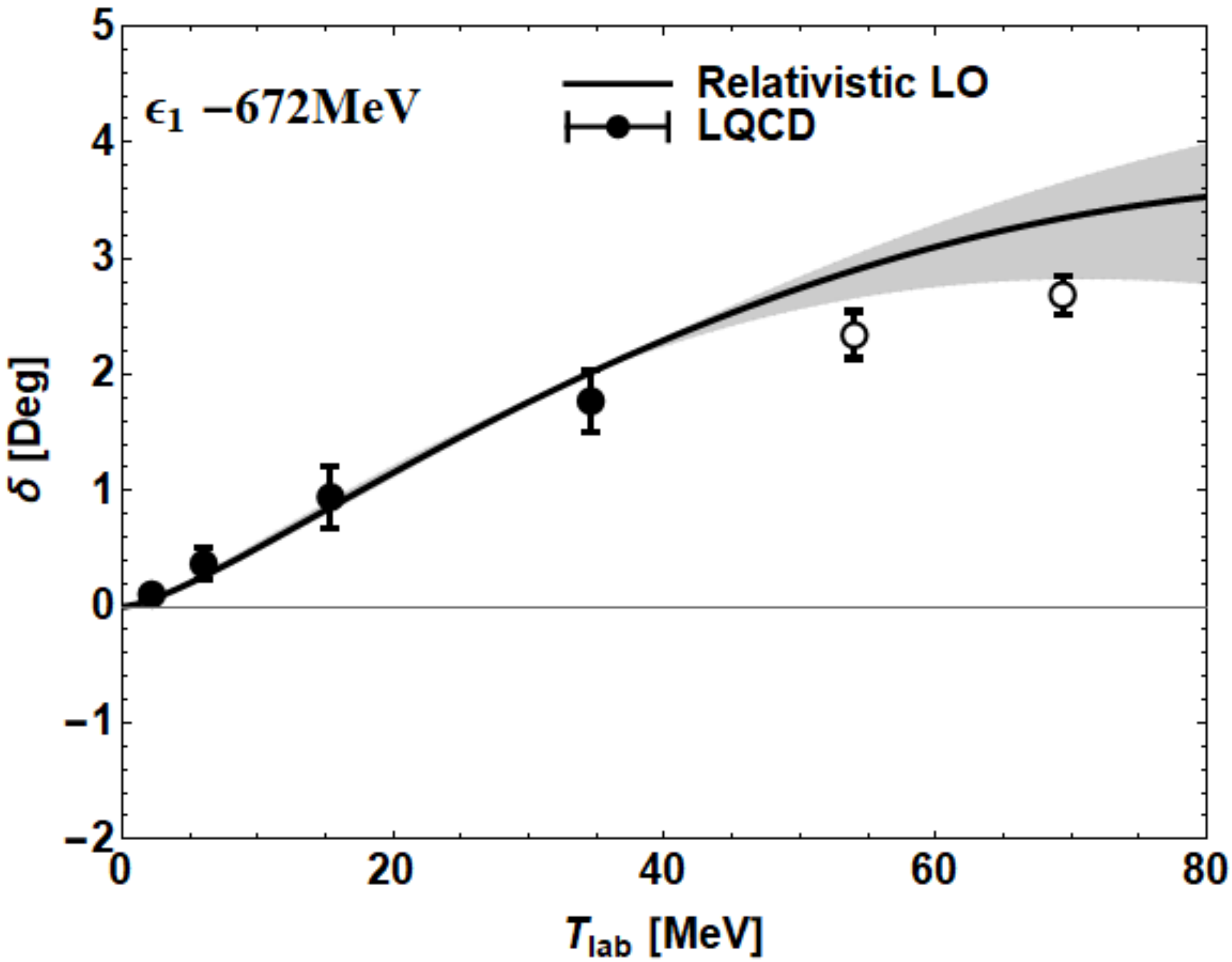}
\includegraphics[scale=0.2]{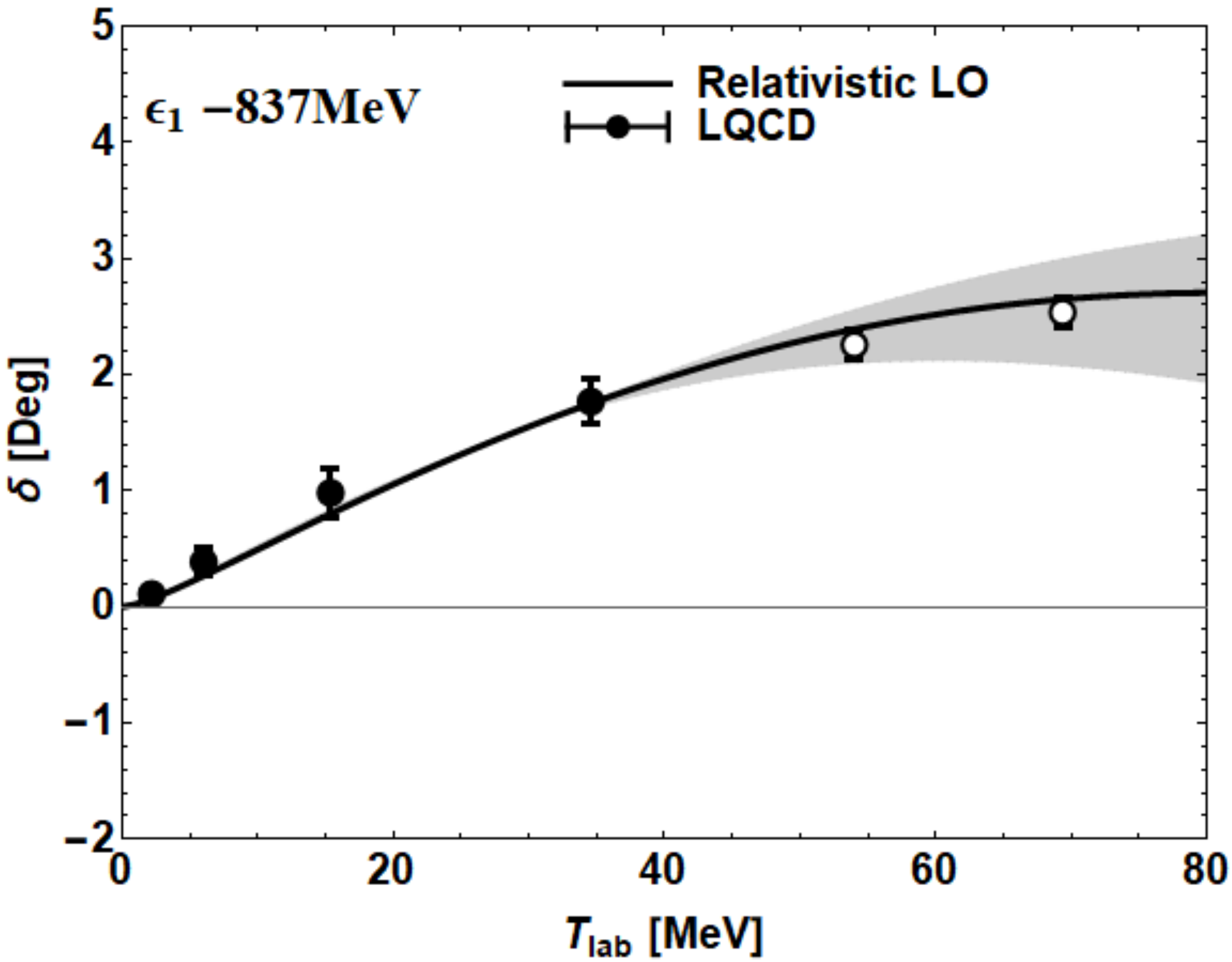}
\includegraphics[scale=0.2]{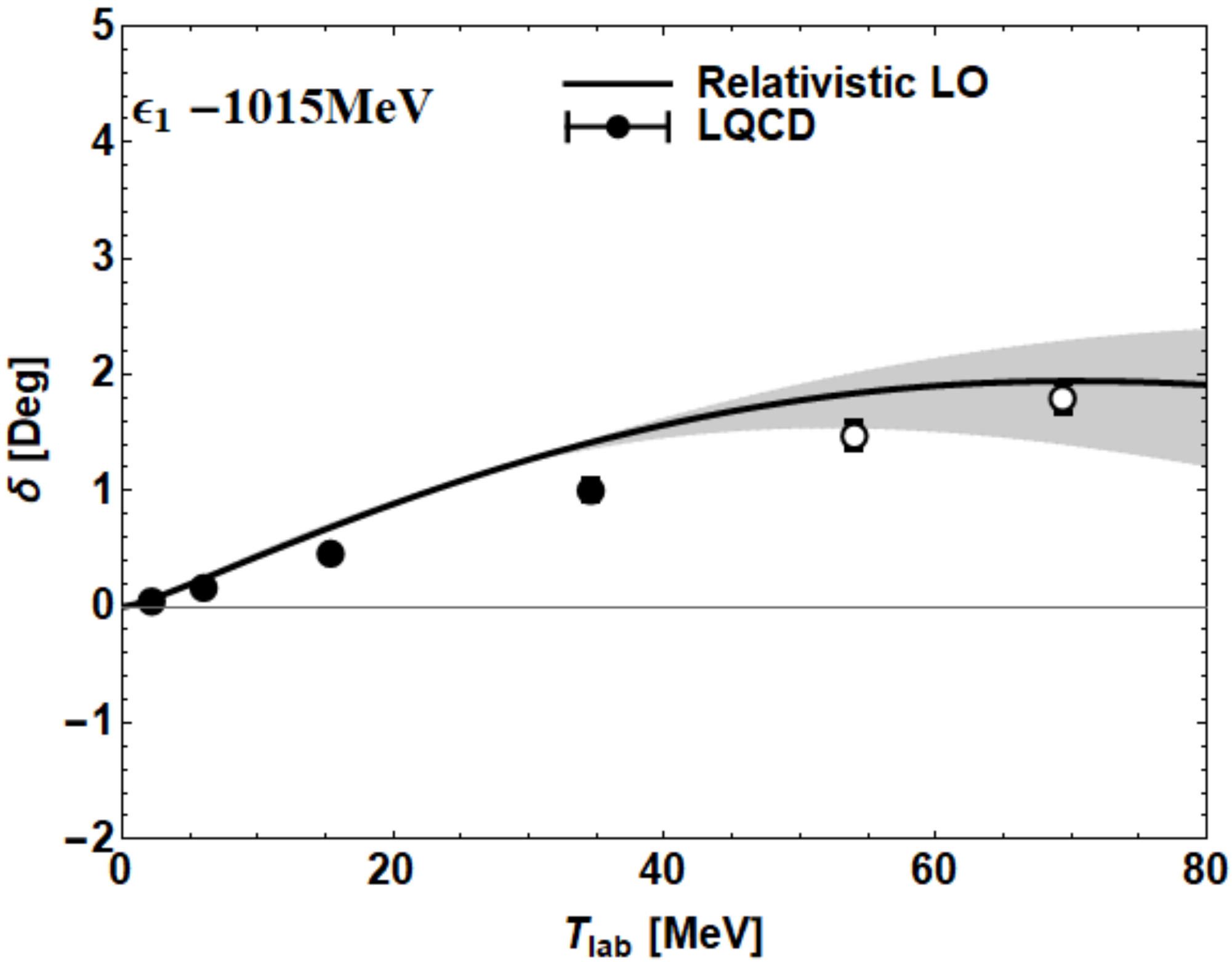}
\includegraphics[scale=0.2]{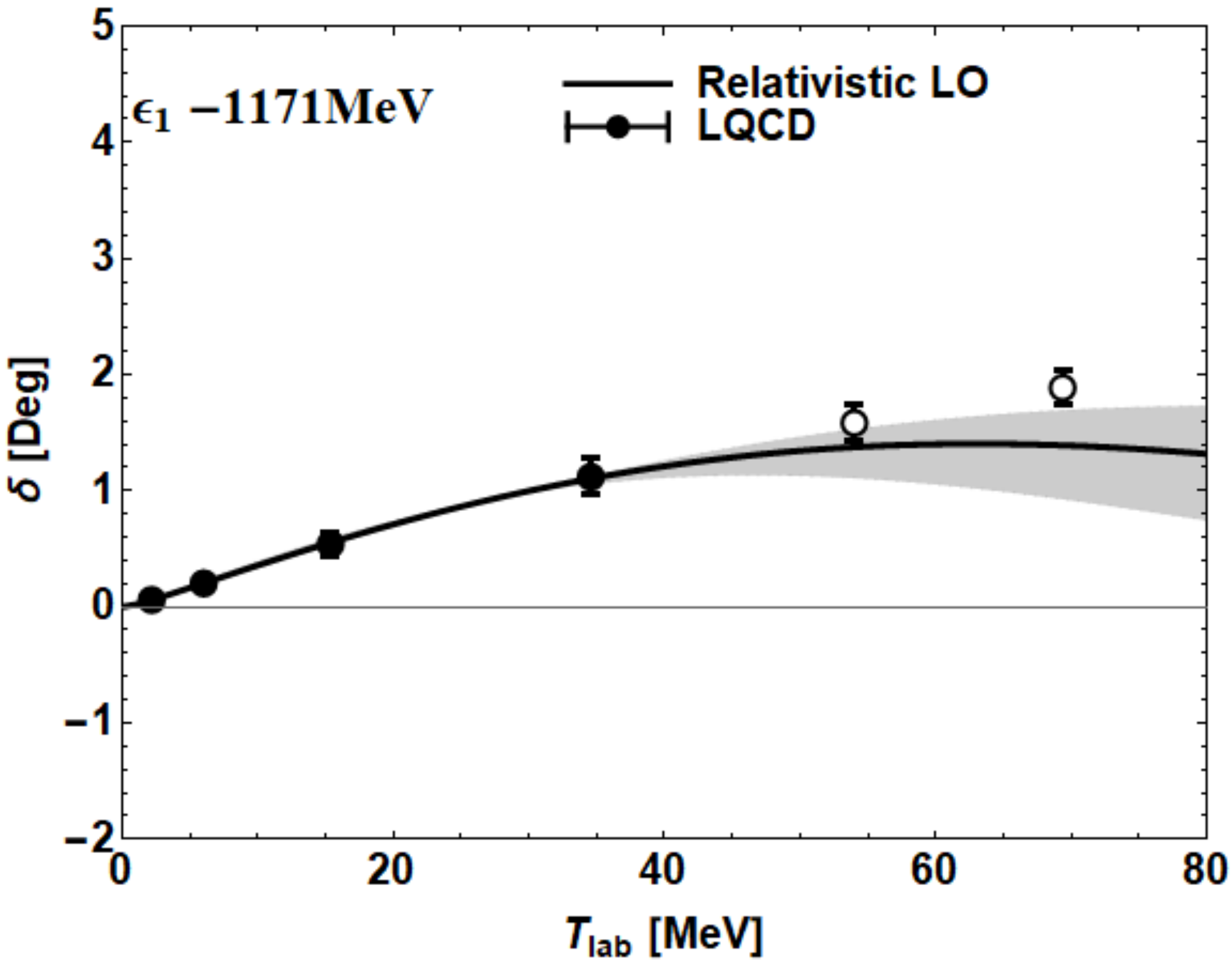}
\caption{ $^3 S_1$ - $^3 D_1$ phase shifts of the lattice QCD simulations with $m_\pi$ = 672, 837, 1015, 1171 MeV in
comparison with the covariant ChEFT fits. The solid  circles represent lattice data fitted, while those empty
circles are not included in the fits. The grey  bands are generated by varying the cutoff from
the optimal value of 360 MeV by $\pm50$ MeV.}\label{fig:E1lat}
\end{figure*}

Next we would like to study the lattice QCD simulations obtained with $m_\pi=672, 837, 1015, 1171$ MeV. As stressed earlier, for such large pion masses, the pion-less ChEFT is more suitable. Therefore, we turn off the one-pion exchanges in solving the Kadyshevsky equation. The results are shown in Figs. 4 and 5, and Tables IV and V. The pion-less version of the covariant ChEFT can fit the four lattice QCD simulations rather well, which is somehow unexpected. Again, the description of the $^1 S_0$ channel is a bit better than that of the  $^3 S_1$ - $^3 D_1 $ coupled channel.  We note that the optimal cutoff for the singlet channel is also different from that for the triplet channel, similar to the pion-full case.

We have checked that it is impossible to fit all the five lattice QCD data with $m_\pi=469,672,837,1015,1171$ MeV simultaneously  by either the pion-full theory
or the pion-less theory. In addition, a naive extrapolation of the pion-less theory to the physical world does not give reasonable phase shifts in comparison with the experimental data.

In principle, these lattice QCD simulations with large pion masses are of no direct relevance to our understanding of the physical world.
Nevertheless, as mentioned earlier, they can be used to test theoretical methods and explore what will happen in
a world with a pion mass different from ours~\cite{Barnea:2013uqa,McIlroy:2017ssf,Hu:2020djy}.

\begin{table}[htpb]
\centering
\caption{ Values of the LECs of the best fit to the lattice QCD simulations with $m_\pi$ = 672, 837, 1015, 1171 MeV for the $^1 S_0$ channel.}\label{tab:1s0lat}
\begin{tabular}{c| c| c | c |c}
 \hline\hline
  $C_{1S0}$  [MeV$^{-2}$] & $\hat{C}_{1S0}$   [MeV$^{-2}$]  & $C_{1S0}^\pi$  [MeV$^{-4}$] & $\hat{C}_{1S0}^{\pi}$  [MeV$^{-4}$] & $\Lambda$ [MeV]\\ [0.3ex]
 \hline
 0.264E-04 & -0.114E-02 & -0.987E-11 & 0.491E-09 & 730\\[0.3ex]
 \hline\hline
\end{tabular}
\end{table}

\begin{table}[htpb]
\centering
\caption{ Values of the LECs of the best fit to the lattice QCD simulations with $m_\pi$ = 672, 837, 1015, 1171 MeV for the $^3 S_1$ - $^3 D_1$ channel.}\label{tab:3s1lat}
\begin{tabular}{c| c| c | c |c}
 \hline\hline
  $C_{3S1}$ [MeV$^{-2}$]  & $\hat{C}_{3S1}$ [MeV$^{-2}$]  & $C_{3S1}^\pi$ [MeV$^{-4}$] & $\hat{C}_{3S1}^{\pi}$  [MeV$^{-4}$] & $\Lambda$ [MeV] \\ [0.3ex]
 \hline
 0.295E-04 & 0.219E-02 & -0.534E-11 & -0.771E-09 & 360 \\[0.3ex]
 \hline\hline
\end{tabular}
\end{table}

\section{Summary and Outlook}
We have studied the lattice QCD simulations of nucleon-nucleon phase shifts together with the experimental data by
the leading order covariant chiral effective field theory. Supplemented with pion-mass dependent low-energy constants,
we showed that the pion-full ChEFT can describe the experimental data and lattice QCD data with $m_\pi = 469$ MeV reasonably well such that chiral extrapolations can be carried out in reasonable confidence, though the description of the $^3 S_1$ - $^3 D_1 $ coupled channel can still be improved. The relative insensitivity of the $^3 S_1$ - $^3 D_1 $ mixing angle on the pion mass remains to be understood. For lattice QCD simulations with $m_\pi$ = 672, 837, 1015, 1171 MeV, the pion-less version of the covariant ChEFT can describe all the four data sets simultaneously, which is a bit unexpected. Here, again, we saw a slightly better performance of the ChEFT in the $^1 S_0 $ channel than that in the  $^3 S_1$ - $^3 D_1 $  coupled channel. The different performance in the singlet and triplet channels can either be attributed to the
fact the leading order covariant ChEFT is insufficient such that higher order studies are necessary or
that there is something that we do not fully understand in the lattice QCD simulations, i.e., the weak dependence of the mixing angle on the pion mass.

Nucleon-nucleon interactions, and more broadly, baryon-baryon interactions play an important role in our understanding of the non-perturbative strong interaction. Lattice QCD simulations are making impressive progress and start to offer new insights on many long-standing issues. To make better
use of such valuable simulation results, their dependence on light quark masses need to be studied in more detail. We have seen tremendous progress in the mesonic and one-baryon sectors in this regard, and hope that the present work can  inspire more works in the two-baryon sector.

\section*{Declaration of competing interest}

The authors declare that they have no known competing financial interests or personal relationships that could have appeared to
influence the work reported in this paper.

\section*{Acknowledgements}

We thank Dr. Takumi Doi and Dr. Takahashi Inoue for providing us the lattice QCD data studied in the present work.
This work was partly supported the National Natural Science Foundation of China (NSFC) under Grants Nos. 11975041, 11735003, and 11961141004.


\bibliography{nn-mpi}

\begin{thebibliography}{46}
\expandafter\ifx\csname natexlab\endcsname\relax\def\natexlab#1{#1}\fi
\expandafter\ifx\csname bibnamefont\endcsname\relax
  \def\bibnamefont#1{#1}\fi
\expandafter\ifx\csname bibfnamefont\endcsname\relax
  \def\bibfnamefont#1{#1}\fi
\expandafter\ifx\csname citenamefont\endcsname\relax
  \def\citenamefont#1{#1}\fi
\expandafter\ifx\csname url\endcsname\relax
  \def\url#1{\texttt{#1}}\fi
\expandafter\ifx\csname urlprefix\endcsname\relax\def\urlprefix{URL }\fi
\providecommand{\bibinfo}[2]{#2}
\providecommand{\eprint}[2][]{\url{#2}}

\bibitem[{\citenamefont{Scherer and Schindler}(2012)}]{Scherer:2012xha}
\bibinfo{author}{\bibfnamefont{S.}~\bibnamefont{Scherer}} \bibnamefont{and}
  \bibinfo{author}{\bibfnamefont{M.~R.} \bibnamefont{Schindler}},
  \emph{\bibinfo{title}{{A Primer for Chiral Perturbation Theory}}}, vol.
  \bibinfo{volume}{830} (\bibinfo{year}{2012}), ISBN
  \bibinfo{isbn}{978-3-642-19253-1}.

\bibitem[{\citenamefont{Bedaque and van Kolck}(2002)}]{Bedaque:2002mn}
\bibinfo{author}{\bibfnamefont{P.~F.} \bibnamefont{Bedaque}} \bibnamefont{and}
  \bibinfo{author}{\bibfnamefont{U.}~\bibnamefont{van Kolck}},
  \bibinfo{journal}{Ann. Rev. Nucl. Part. Sci.} \textbf{\bibinfo{volume}{52}},
  \bibinfo{pages}{339} (\bibinfo{year}{2002}), \eprint{nucl-th/0203055}.

\bibitem[{\citenamefont{Epelbaum et~al.}(2009)\citenamefont{Epelbaum, Hammer,
  and Meissner}}]{Epelbaum:2008ga}
\bibinfo{author}{\bibfnamefont{E.}~\bibnamefont{Epelbaum}},
  \bibinfo{author}{\bibfnamefont{H.-W.} \bibnamefont{Hammer}},
  \bibnamefont{and} \bibinfo{author}{\bibfnamefont{U.-G.}
  \bibnamefont{Meissner}}, \bibinfo{journal}{Rev. Mod. Phys.}
  \textbf{\bibinfo{volume}{81}}, \bibinfo{pages}{1773} (\bibinfo{year}{2009}),
  \eprint{0811.1338}.

\bibitem[{\citenamefont{Machleidt and Entem}(2011)}]{Machleidt:2011zz}
\bibinfo{author}{\bibfnamefont{R.}~\bibnamefont{Machleidt}} \bibnamefont{and}
  \bibinfo{author}{\bibfnamefont{D.~R.} \bibnamefont{Entem}},
  \bibinfo{journal}{Phys. Rept.} \textbf{\bibinfo{volume}{503}},
  \bibinfo{pages}{1} (\bibinfo{year}{2011}), \eprint{1105.2919}.

\bibitem[{\citenamefont{Weinberg}(1990)}]{Weinberg:1990rz}
\bibinfo{author}{\bibfnamefont{S.}~\bibnamefont{Weinberg}},
  \bibinfo{journal}{Phys. Lett.} \textbf{\bibinfo{volume}{B251}},
  \bibinfo{pages}{288} (\bibinfo{year}{1990}).

\bibitem[{\citenamefont{Weinberg}(1991)}]{Weinberg:1991um}
\bibinfo{author}{\bibfnamefont{S.}~\bibnamefont{Weinberg}},
  \bibinfo{journal}{Nucl. Phys.} \textbf{\bibinfo{volume}{B363}},
  \bibinfo{pages}{3} (\bibinfo{year}{1991}).

\bibitem[{\citenamefont{Entem and Machleidt}(2003)}]{Entem:2003ft}
\bibinfo{author}{\bibfnamefont{D.~R.} \bibnamefont{Entem}} \bibnamefont{and}
  \bibinfo{author}{\bibfnamefont{R.}~\bibnamefont{Machleidt}},
  \bibinfo{journal}{Phys. Rev.} \textbf{\bibinfo{volume}{C68}},
  \bibinfo{pages}{041001} (\bibinfo{year}{2003}), \eprint{nucl-th/0304018}.

\bibitem[{\citenamefont{Epelbaum et~al.}(2005)\citenamefont{Epelbaum, Glockle,
  and Meissner}}]{Epelbaum:2004fk}
\bibinfo{author}{\bibfnamefont{E.}~\bibnamefont{Epelbaum}},
  \bibinfo{author}{\bibfnamefont{W.}~\bibnamefont{Glockle}}, \bibnamefont{and}
  \bibinfo{author}{\bibfnamefont{U.-G.} \bibnamefont{Meissner}},
  \bibinfo{journal}{Nucl. Phys. A} \textbf{\bibinfo{volume}{747}},
  \bibinfo{pages}{362} (\bibinfo{year}{2005}), \eprint{nucl-th/0405048}.

\bibitem[{\citenamefont{Entem et~al.}(2017)\citenamefont{Entem, Machleidt, and
  Nosyk}}]{Entem:2017gor}
\bibinfo{author}{\bibfnamefont{D.}~\bibnamefont{Entem}},
  \bibinfo{author}{\bibfnamefont{R.}~\bibnamefont{Machleidt}},
  \bibnamefont{and} \bibinfo{author}{\bibfnamefont{Y.}~\bibnamefont{Nosyk}},
  \bibinfo{journal}{Phys. Rev. C} \textbf{\bibinfo{volume}{96}},
  \bibinfo{pages}{024004} (\bibinfo{year}{2017}), \eprint{1703.05454}.

\bibitem[{\citenamefont{Reinert et~al.}(2018)\citenamefont{Reinert, Krebs, and
  Epelbaum}}]{Reinert:2017usi}
\bibinfo{author}{\bibfnamefont{P.}~\bibnamefont{Reinert}},
  \bibinfo{author}{\bibfnamefont{H.}~\bibnamefont{Krebs}}, \bibnamefont{and}
  \bibinfo{author}{\bibfnamefont{E.}~\bibnamefont{Epelbaum}},
  \bibinfo{journal}{Eur. Phys. J. A} \textbf{\bibinfo{volume}{54}},
  \bibinfo{pages}{86} (\bibinfo{year}{2018}), \eprint{1711.08821}.

\bibitem[{\citenamefont{Hammer et~al.}(2019)\citenamefont{Hammer, König, and
  van Kolck}}]{Hammer:2019poc}
\bibinfo{author}{\bibfnamefont{H.-W.} \bibnamefont{Hammer}},
  \bibinfo{author}{\bibfnamefont{S.}~\bibnamefont{König}}, \bibnamefont{and}
  \bibinfo{author}{\bibfnamefont{U.}~\bibnamefont{van Kolck}}
  (\bibinfo{year}{2019}), \eprint{1906.12122}.

\bibitem[{\citenamefont{Epelbaum et~al.}(2020)\citenamefont{Epelbaum, Krebs,
  and Reinert}}]{Epelbaum:2019kcf}
\bibinfo{author}{\bibfnamefont{E.}~\bibnamefont{Epelbaum}},
  \bibinfo{author}{\bibfnamefont{H.}~\bibnamefont{Krebs}}, \bibnamefont{and}
  \bibinfo{author}{\bibfnamefont{P.}~\bibnamefont{Reinert}},
  \bibinfo{journal}{Front. in Phys.} \textbf{\bibinfo{volume}{8}},
  \bibinfo{pages}{98} (\bibinfo{year}{2020}), \eprint{1911.11875}.

\bibitem[{\citenamefont{Rodriguez~Entem
  et~al.}(2020)\citenamefont{Rodriguez~Entem, Machleidt, and
  Nosyk}}]{RodriguezEntem:2020jgp}
\bibinfo{author}{\bibfnamefont{D.}~\bibnamefont{Rodriguez~Entem}},
  \bibinfo{author}{\bibfnamefont{R.}~\bibnamefont{Machleidt}},
  \bibnamefont{and} \bibinfo{author}{\bibfnamefont{Y.}~\bibnamefont{Nosyk}},
  \bibinfo{journal}{Front. in Phys.} \textbf{\bibinfo{volume}{8}},
  \bibinfo{pages}{57} (\bibinfo{year}{2020}).

\bibitem[{\citenamefont{Aoki and Doi}(2020)}]{Aoki:2020bew}
\bibinfo{author}{\bibfnamefont{S.}~\bibnamefont{Aoki}} \bibnamefont{and}
  \bibinfo{author}{\bibfnamefont{T.}~\bibnamefont{Doi}} (\bibinfo{year}{2020}),
  \eprint{2003.10730}.

\bibitem[{\citenamefont{Ishii et~al.}(2007)\citenamefont{Ishii, Aoki, and
  Hatsuda}}]{Ishii:2006ec}
\bibinfo{author}{\bibfnamefont{N.}~\bibnamefont{Ishii}},
  \bibinfo{author}{\bibfnamefont{S.}~\bibnamefont{Aoki}}, \bibnamefont{and}
  \bibinfo{author}{\bibfnamefont{T.}~\bibnamefont{Hatsuda}},
  \bibinfo{journal}{Phys. Rev. Lett.} \textbf{\bibinfo{volume}{99}},
  \bibinfo{pages}{022001} (\bibinfo{year}{2007}), \eprint{nucl-th/0611096}.

\bibitem[{\citenamefont{Barnea et~al.}(2015)\citenamefont{Barnea, Contessi,
  Gazit, Pederiva, and van Kolck}}]{Barnea:2013uqa}
\bibinfo{author}{\bibfnamefont{N.}~\bibnamefont{Barnea}},
  \bibinfo{author}{\bibfnamefont{L.}~\bibnamefont{Contessi}},
  \bibinfo{author}{\bibfnamefont{D.}~\bibnamefont{Gazit}},
  \bibinfo{author}{\bibfnamefont{F.}~\bibnamefont{Pederiva}}, \bibnamefont{and}
  \bibinfo{author}{\bibfnamefont{U.}~\bibnamefont{van Kolck}},
  \bibinfo{journal}{Phys. Rev. Lett.} \textbf{\bibinfo{volume}{114}},
  \bibinfo{pages}{052501} (\bibinfo{year}{2015}), \eprint{1311.4966}.

\bibitem[{\citenamefont{McIlroy et~al.}(2018)\citenamefont{McIlroy, Barbieri,
  Inoue, Doi, and Hatsuda}}]{McIlroy:2017ssf}
\bibinfo{author}{\bibfnamefont{C.}~\bibnamefont{McIlroy}},
  \bibinfo{author}{\bibfnamefont{C.}~\bibnamefont{Barbieri}},
  \bibinfo{author}{\bibfnamefont{T.}~\bibnamefont{Inoue}},
  \bibinfo{author}{\bibfnamefont{T.}~\bibnamefont{Doi}}, \bibnamefont{and}
  \bibinfo{author}{\bibfnamefont{T.}~\bibnamefont{Hatsuda}},
  \bibinfo{journal}{Phys. Rev. C} \textbf{\bibinfo{volume}{97}},
  \bibinfo{pages}{021303} (\bibinfo{year}{2018}), \eprint{1701.02607}.

\bibitem[{\citenamefont{Hu et~al.}(2020)\citenamefont{Hu, Zhang, Shen, and
  Toki}}]{Hu:2020djy}
\bibinfo{author}{\bibfnamefont{J.}~\bibnamefont{Hu}},
  \bibinfo{author}{\bibfnamefont{Y.}~\bibnamefont{Zhang}},
  \bibinfo{author}{\bibfnamefont{H.}~\bibnamefont{Shen}}, \bibnamefont{and}
  \bibinfo{author}{\bibfnamefont{H.}~\bibnamefont{Toki}}
  (\bibinfo{year}{2020}), \eprint{2003.08008}.

\bibitem[{\citenamefont{Ren et~al.}(2018{\natexlab{a}})\citenamefont{Ren, Ling,
  and Geng}}]{Ling:2017jyz}
\bibinfo{author}{\bibfnamefont{X.-L.} \bibnamefont{Ren}},
  \bibinfo{author}{\bibfnamefont{X.-Z.} \bibnamefont{Ling}}, \bibnamefont{and}
  \bibinfo{author}{\bibfnamefont{L.-S.} \bibnamefont{Geng}},
  \bibinfo{journal}{Phys. Lett. B} \textbf{\bibinfo{volume}{783}},
  \bibinfo{pages}{7} (\bibinfo{year}{2018}{\natexlab{a}}), \eprint{1710.07164}.

\bibitem[{\citenamefont{Xiao et~al.}(2018)\citenamefont{Xiao, Ren, Lu, Geng,
  and Meißner}}]{Xiao:2018rvd}
\bibinfo{author}{\bibfnamefont{Y.}~\bibnamefont{Xiao}},
  \bibinfo{author}{\bibfnamefont{X.-L.} \bibnamefont{Ren}},
  \bibinfo{author}{\bibfnamefont{J.-X.} \bibnamefont{Lu}},
  \bibinfo{author}{\bibfnamefont{L.-S.} \bibnamefont{Geng}}, \bibnamefont{and}
  \bibinfo{author}{\bibfnamefont{U.-G.} \bibnamefont{Meißner}},
  \bibinfo{journal}{Eur. Phys. J. C} \textbf{\bibinfo{volume}{78}},
  \bibinfo{pages}{489} (\bibinfo{year}{2018}), \eprint{1803.04251}.

\bibitem[{\citenamefont{Eliyahu et~al.}(2019)\citenamefont{Eliyahu, Bazak, and
  Barnea}}]{Eliyahu:2019nkz}
\bibinfo{author}{\bibfnamefont{M.}~\bibnamefont{Eliyahu}},
  \bibinfo{author}{\bibfnamefont{B.}~\bibnamefont{Bazak}}, \bibnamefont{and}
  \bibinfo{author}{\bibfnamefont{N.}~\bibnamefont{Barnea}}
  (\bibinfo{year}{2019}), \eprint{1912.07017}.

\bibitem[{\citenamefont{Song et~al.}(2018)\citenamefont{Song, Li, and
  Geng}}]{Song:2018qqm}
\bibinfo{author}{\bibfnamefont{J.}~\bibnamefont{Song}},
  \bibinfo{author}{\bibfnamefont{K.-W.} \bibnamefont{Li}}, \bibnamefont{and}
  \bibinfo{author}{\bibfnamefont{L.-S.} \bibnamefont{Geng}},
  \bibinfo{journal}{Phys.\ Rev.\ C} \textbf{\bibinfo{volume}{97}},
  \bibinfo{pages}{065201} (\bibinfo{year}{2018}), \eprint{1802.04433}.

\bibitem[{\citenamefont{Li et~al.}(2018{\natexlab{a}})\citenamefont{Li, Hyodo,
  and Geng}}]{Li:2018tbt}
\bibinfo{author}{\bibfnamefont{K.-W.} \bibnamefont{Li}},
  \bibinfo{author}{\bibfnamefont{T.}~\bibnamefont{Hyodo}}, \bibnamefont{and}
  \bibinfo{author}{\bibfnamefont{L.-S.} \bibnamefont{Geng}},
  \bibinfo{journal}{Phys.\ Rev.\ C} \textbf{\bibinfo{volume}{98}},
  \bibinfo{pages}{065203} (\bibinfo{year}{2018}{\natexlab{a}}),
  \eprint{1809.03199}.

\bibitem[{\citenamefont{Nemura et~al.}(2018)}]{Nemura:2017vjc}
\bibinfo{author}{\bibfnamefont{H.}~\bibnamefont{Nemura}} \bibnamefont{et~al.},
  \bibinfo{journal}{EPJ Web Conf.} \textbf{\bibinfo{volume}{175}},
  \bibinfo{pages}{05030} (\bibinfo{year}{2018}), \eprint{1711.07003}.

\bibitem[{\citenamefont{Miyamoto}(2016)}]{Miyamoto:2016hqo}
\bibinfo{author}{\bibfnamefont{T.}~\bibnamefont{Miyamoto}}
  (\bibinfo{collaboration}{HAL QCD}), \bibinfo{journal}{PoS}
  \textbf{\bibinfo{volume}{LATTICE2015}}, \bibinfo{pages}{090}
  (\bibinfo{year}{2016}), \eprint{1602.07797}.

\bibitem[{\citenamefont{Beane et~al.}(2007)\citenamefont{Beane, Bedaque, Luu,
  Orginos, Pallante, Parreno, and Savage}}]{Beane:2006gf}
\bibinfo{author}{\bibfnamefont{S.~R.} \bibnamefont{Beane}},
  \bibinfo{author}{\bibfnamefont{P.~F.} \bibnamefont{Bedaque}},
  \bibinfo{author}{\bibfnamefont{T.~C.} \bibnamefont{Luu}},
  \bibinfo{author}{\bibfnamefont{K.}~\bibnamefont{Orginos}},
  \bibinfo{author}{\bibfnamefont{E.}~\bibnamefont{Pallante}},
  \bibinfo{author}{\bibfnamefont{A.}~\bibnamefont{Parreno}}, \bibnamefont{and}
  \bibinfo{author}{\bibfnamefont{M.~J.} \bibnamefont{Savage}}
  (\bibinfo{collaboration}{NPLQCD}), \bibinfo{journal}{Nucl. Phys. A}
  \textbf{\bibinfo{volume}{794}}, \bibinfo{pages}{62} (\bibinfo{year}{2007}),
  \eprint{hep-lat/0612026}.

\bibitem[{\citenamefont{Sasaki et~al.}(2018)\citenamefont{Sasaki, Aoki, Doi,
  Gongyo, Hatsuda, Ikeda, Inoue, Iritani, Ishii, and
  Miyamoto}}]{Sasaki:2018mzh}
\bibinfo{author}{\bibfnamefont{K.}~\bibnamefont{Sasaki}},
  \bibinfo{author}{\bibfnamefont{S.}~\bibnamefont{Aoki}},
  \bibinfo{author}{\bibfnamefont{T.}~\bibnamefont{Doi}},
  \bibinfo{author}{\bibfnamefont{S.}~\bibnamefont{Gongyo}},
  \bibinfo{author}{\bibfnamefont{T.}~\bibnamefont{Hatsuda}},
  \bibinfo{author}{\bibfnamefont{Y.}~\bibnamefont{Ikeda}},
  \bibinfo{author}{\bibfnamefont{T.}~\bibnamefont{Inoue}},
  \bibinfo{author}{\bibfnamefont{T.}~\bibnamefont{Iritani}},
  \bibinfo{author}{\bibfnamefont{N.}~\bibnamefont{Ishii}}, \bibnamefont{and}
  \bibinfo{author}{\bibfnamefont{T.}~\bibnamefont{Miyamoto}}
  (\bibinfo{collaboration}{HAL QCD}), \bibinfo{journal}{EPJ Web Conf.}
  \textbf{\bibinfo{volume}{175}}, \bibinfo{pages}{05010}
  (\bibinfo{year}{2018}).

\bibitem[{\citenamefont{Inoue et~al.}(2012)\citenamefont{Inoue, Aoki, Doi,
  Hatsuda, Ikeda, Ishii, Murano, Nemura, and Sasaki}}]{Inoue:2011ai}
\bibinfo{author}{\bibfnamefont{T.}~\bibnamefont{Inoue}},
  \bibinfo{author}{\bibfnamefont{S.}~\bibnamefont{Aoki}},
  \bibinfo{author}{\bibfnamefont{T.}~\bibnamefont{Doi}},
  \bibinfo{author}{\bibfnamefont{T.}~\bibnamefont{Hatsuda}},
  \bibinfo{author}{\bibfnamefont{Y.}~\bibnamefont{Ikeda}},
  \bibinfo{author}{\bibfnamefont{N.}~\bibnamefont{Ishii}},
  \bibinfo{author}{\bibfnamefont{K.}~\bibnamefont{Murano}},
  \bibinfo{author}{\bibfnamefont{H.}~\bibnamefont{Nemura}}, \bibnamefont{and}
  \bibinfo{author}{\bibfnamefont{K.}~\bibnamefont{Sasaki}}
  (\bibinfo{collaboration}{HAL QCD}), \bibinfo{journal}{Nucl. Phys. A}
  \textbf{\bibinfo{volume}{881}}, \bibinfo{pages}{28} (\bibinfo{year}{2012}),
  \eprint{1112.5926}.

\bibitem[{\citenamefont{Baru et~al.}(2016)\citenamefont{Baru, Epelbaum, and
  Filin}}]{Baru:2016evv}
\bibinfo{author}{\bibfnamefont{V.}~\bibnamefont{Baru}},
  \bibinfo{author}{\bibfnamefont{E.}~\bibnamefont{Epelbaum}}, \bibnamefont{and}
  \bibinfo{author}{\bibfnamefont{A.}~\bibnamefont{Filin}},
  \bibinfo{journal}{Phys. Rev. C} \textbf{\bibinfo{volume}{94}},
  \bibinfo{pages}{014001} (\bibinfo{year}{2016}), \eprint{1604.02551}.

\bibitem[{\citenamefont{Ren et~al.}(2018{\natexlab{b}})\citenamefont{Ren, Li,
  Geng, Long, Ring, and Meng}}]{Ren:2016jna}
\bibinfo{author}{\bibfnamefont{X.-L.} \bibnamefont{Ren}},
  \bibinfo{author}{\bibfnamefont{K.-W.} \bibnamefont{Li}},
  \bibinfo{author}{\bibfnamefont{L.-S.} \bibnamefont{Geng}},
  \bibinfo{author}{\bibfnamefont{B.-W.} \bibnamefont{Long}},
  \bibinfo{author}{\bibfnamefont{P.}~\bibnamefont{Ring}}, \bibnamefont{and}
  \bibinfo{author}{\bibfnamefont{J.}~\bibnamefont{Meng}},
  \bibinfo{journal}{Chin. Phys.} \textbf{\bibinfo{volume}{C42}},
  \bibinfo{pages}{014103} (\bibinfo{year}{2018}{\natexlab{b}}),
  \eprint{1611.08475}.

\bibitem[{\citenamefont{Li et~al.}(2018{\natexlab{b}})\citenamefont{Li, Ren,
  Geng, and Long}}]{Li:2016mln}
\bibinfo{author}{\bibfnamefont{K.-W.} \bibnamefont{Li}},
  \bibinfo{author}{\bibfnamefont{X.-L.} \bibnamefont{Ren}},
  \bibinfo{author}{\bibfnamefont{L.-S.} \bibnamefont{Geng}}, \bibnamefont{and}
  \bibinfo{author}{\bibfnamefont{B.-W.} \bibnamefont{Long}},
  \bibinfo{journal}{Chin. Phys.} \textbf{\bibinfo{volume}{C42}},
  \bibinfo{pages}{014105} (\bibinfo{year}{2018}{\natexlab{b}}),
  \eprint{1612.08482}.

\bibitem[{\citenamefont{Xiao et~al.}(2019)\citenamefont{Xiao, Geng, and
  Ren}}]{Xiao:2018jot}
\bibinfo{author}{\bibfnamefont{Y.}~\bibnamefont{Xiao}},
  \bibinfo{author}{\bibfnamefont{L.-S.} \bibnamefont{Geng}}, \bibnamefont{and}
  \bibinfo{author}{\bibfnamefont{X.-L.} \bibnamefont{Ren}},
  \bibinfo{journal}{Phys. Rev.} \textbf{\bibinfo{volume}{C99}},
  \bibinfo{pages}{024004} (\bibinfo{year}{2019}), \eprint{1812.03005}.

\bibitem[{\citenamefont{Wang et~al.}(2020)\citenamefont{Wang, Geng, and
  Long}}]{wang:2020myr}
\bibinfo{author}{\bibfnamefont{C.-X.} \bibnamefont{Wang}},
  \bibinfo{author}{\bibfnamefont{L.-S.} \bibnamefont{Geng}}, \bibnamefont{and}
  \bibinfo{author}{\bibfnamefont{B.}~\bibnamefont{Long}}
  (\bibinfo{year}{2020}), \eprint{2001.08483}.

\bibitem[{\citenamefont{Haidenbauer and Krein}(2018)}]{Haidenbauer:2017dua}
\bibinfo{author}{\bibfnamefont{J.}~\bibnamefont{Haidenbauer}} \bibnamefont{and}
  \bibinfo{author}{\bibfnamefont{G.}~\bibnamefont{Krein}},
  \bibinfo{journal}{Eur. Phys. J. A} \textbf{\bibinfo{volume}{54}},
  \bibinfo{pages}{199} (\bibinfo{year}{2018}), \eprint{1711.06470}.

\bibitem[{\citenamefont{Beane and Savage}(2003{\natexlab{a}})}]{Beane:2002xf}
\bibinfo{author}{\bibfnamefont{S.~R.} \bibnamefont{Beane}} \bibnamefont{and}
  \bibinfo{author}{\bibfnamefont{M.~J.} \bibnamefont{Savage}},
  \bibinfo{journal}{Nucl. Phys. A} \textbf{\bibinfo{volume}{717}},
  \bibinfo{pages}{91} (\bibinfo{year}{2003}{\natexlab{a}}),
  \eprint{nucl-th/0208021}.

\bibitem[{\citenamefont{Beane and Savage}(2003{\natexlab{b}})}]{Beane:2002vs}
\bibinfo{author}{\bibfnamefont{S.~R.} \bibnamefont{Beane}} \bibnamefont{and}
  \bibinfo{author}{\bibfnamefont{M.~J.} \bibnamefont{Savage}},
  \bibinfo{journal}{Nucl. Phys. A} \textbf{\bibinfo{volume}{713}},
  \bibinfo{pages}{148} (\bibinfo{year}{2003}{\natexlab{b}}),
  \eprint{hep-ph/0206113}.

\bibitem[{\citenamefont{Kadyshevsky}(1968)}]{Kadyshevsky:1967rs}
\bibinfo{author}{\bibfnamefont{V.~G.} \bibnamefont{Kadyshevsky}},
  \bibinfo{journal}{Nucl. Phys.} \textbf{\bibinfo{volume}{B6}},
  \bibinfo{pages}{125} (\bibinfo{year}{1968}).

\bibitem[{\citenamefont{Iritani et~al.}(2017)\citenamefont{Iritani, Aoki, Doi,
  Hatsuda, Ikeda, Inoue, Ishii, Nemura, and Sasaki}}]{Iritani:2017rlk}
\bibinfo{author}{\bibfnamefont{T.}~\bibnamefont{Iritani}},
  \bibinfo{author}{\bibfnamefont{S.}~\bibnamefont{Aoki}},
  \bibinfo{author}{\bibfnamefont{T.}~\bibnamefont{Doi}},
  \bibinfo{author}{\bibfnamefont{T.}~\bibnamefont{Hatsuda}},
  \bibinfo{author}{\bibfnamefont{Y.}~\bibnamefont{Ikeda}},
  \bibinfo{author}{\bibfnamefont{T.}~\bibnamefont{Inoue}},
  \bibinfo{author}{\bibfnamefont{N.}~\bibnamefont{Ishii}},
  \bibinfo{author}{\bibfnamefont{H.}~\bibnamefont{Nemura}}, \bibnamefont{and}
  \bibinfo{author}{\bibfnamefont{K.}~\bibnamefont{Sasaki}},
  \bibinfo{journal}{Phys. Rev. D} \textbf{\bibinfo{volume}{96}},
  \bibinfo{pages}{034521} (\bibinfo{year}{2017}), \eprint{1703.07210}.

\bibitem[{\citenamefont{Beane et~al.}(2017)}]{Beane:2017edf}
\bibinfo{author}{\bibfnamefont{S.~R.} \bibnamefont{Beane}} \bibnamefont{et~al.}
  (\bibinfo{year}{2017}), \eprint{1705.09239}.

\bibitem[{\citenamefont{Yamazaki and Kuramashi}(2017)}]{Yamazaki:2017gjl}
\bibinfo{author}{\bibfnamefont{T.}~\bibnamefont{Yamazaki}} \bibnamefont{and}
  \bibinfo{author}{\bibfnamefont{Y.}~\bibnamefont{Kuramashi}},
  \bibinfo{journal}{Phys. Rev. D} \textbf{\bibinfo{volume}{96}},
  \bibinfo{pages}{114511} (\bibinfo{year}{2017}), \eprint{1709.09779}.

\bibitem[{\citenamefont{Zhou et~al.}(2014)\citenamefont{Zhou, Ren, Chen, and
  Geng}}]{Zhou:2014ila}
\bibinfo{author}{\bibfnamefont{Y.}~\bibnamefont{Zhou}},
  \bibinfo{author}{\bibfnamefont{X.-L.} \bibnamefont{Ren}},
  \bibinfo{author}{\bibfnamefont{H.-X.} \bibnamefont{Chen}}, \bibnamefont{and}
  \bibinfo{author}{\bibfnamefont{L.-S.} \bibnamefont{Geng}},
  \bibinfo{journal}{Phys. Rev. D} \textbf{\bibinfo{volume}{90}},
  \bibinfo{pages}{014020} (\bibinfo{year}{2014}), \eprint{1404.6847}.

\bibitem[{\citenamefont{Soto and Tarrus}(2012)}]{Soto:2011tb}
\bibinfo{author}{\bibfnamefont{J.}~\bibnamefont{Soto}} \bibnamefont{and}
  \bibinfo{author}{\bibfnamefont{J.}~\bibnamefont{Tarrus}},
  \bibinfo{journal}{Phys. Rev. C} \textbf{\bibinfo{volume}{85}},
  \bibinfo{pages}{044001} (\bibinfo{year}{2012}), \eprint{1112.4426}.

\bibitem[{\citenamefont{Chen et~al.}(2012)\citenamefont{Chen, Lee, Liu, and
  Liu}}]{Chen:2010yt}
\bibinfo{author}{\bibfnamefont{J.-W.} \bibnamefont{Chen}},
  \bibinfo{author}{\bibfnamefont{T.-K.} \bibnamefont{Lee}},
  \bibinfo{author}{\bibfnamefont{C.-P.} \bibnamefont{Liu}}, \bibnamefont{and}
  \bibinfo{author}{\bibfnamefont{Y.-S.} \bibnamefont{Liu}},
  \bibinfo{journal}{Phys. Rev. C} \textbf{\bibinfo{volume}{86}},
  \bibinfo{pages}{054001} (\bibinfo{year}{2012}), \eprint{1012.0453}.

\bibitem[{\citenamefont{Beane et~al.}(2002)\citenamefont{Beane, Bedaque,
  Savage, and van Kolck}}]{Beane:2001bc}
\bibinfo{author}{\bibfnamefont{S.}~\bibnamefont{Beane}},
  \bibinfo{author}{\bibfnamefont{P.~F.} \bibnamefont{Bedaque}},
  \bibinfo{author}{\bibfnamefont{M.}~\bibnamefont{Savage}}, \bibnamefont{and}
  \bibinfo{author}{\bibfnamefont{U.}~\bibnamefont{van Kolck}},
  \bibinfo{journal}{Nucl. Phys. A} \textbf{\bibinfo{volume}{700}},
  \bibinfo{pages}{377} (\bibinfo{year}{2002}), \eprint{nucl-th/0104030}.

\bibitem[{\citenamefont{Epelbaum et~al.}(2002)\citenamefont{Epelbaum, Meissner,
  and Gloeckle}}]{Epelbaum:2002gk}
\bibinfo{author}{\bibfnamefont{E.}~\bibnamefont{Epelbaum}},
  \bibinfo{author}{\bibfnamefont{U.-G.} \bibnamefont{Meissner}},
  \bibnamefont{and} \bibinfo{author}{\bibfnamefont{W.}~\bibnamefont{Gloeckle}}
  (\bibinfo{year}{2002}), \eprint{nucl-th/0208040}.

\bibitem[{\citenamefont{Epelbaum et~al.}(2003)\citenamefont{Epelbaum, Meissner,
  and Gloeckle}}]{Epelbaum:2002gb}
\bibinfo{author}{\bibfnamefont{E.}~\bibnamefont{Epelbaum}},
  \bibinfo{author}{\bibfnamefont{U.-G.} \bibnamefont{Meissner}},
  \bibnamefont{and} \bibinfo{author}{\bibfnamefont{W.}~\bibnamefont{Gloeckle}},
  \bibinfo{journal}{Nucl. Phys. A} \textbf{\bibinfo{volume}{714}},
  \bibinfo{pages}{535} (\bibinfo{year}{2003}), \eprint{nucl-th/0207089}.

\end{thebibliography}
\end{document}